\documentclass[]{jfm}

\usepackage{graphicx}
\usepackage{graphics}
\usepackage[sort]{natbib}
\usepackage{amsmath}
\usepackage{times}
\usepackage{color}
\usepackage{url}

\ifCUPmtlplainloaded \else
  \checkfont{eurm10}
  \iffontfound
    \IfFileExists{upmath.sty}
      {\typeout{^^JFound AMS Euler Roman fonts on the system,
                   using the 'upmath' package.^^J}%
       \usepackage{upmath}}
      {\typeout{^^JFound AMS Euler Roman fonts on the system, but you
                   dont seem to have the}%
       \typeout{'upmath' package installed. JFM.cls can take advantage
                 of these fonts,^^Jif you use 'upmath' package.^^J}%
      }
  \else
  \fi
\fi

\ifCUPmtlplainloaded \else
  \checkfont{msam10}
  \iffontfound
    \IfFileExists{amssymb.sty}
      {\typeout{^^JFound AMS Symbol fonts on the system, using the
                'amssymb' package.^^J}%
       \usepackage{amssymb}%
         
       \let\ge=\geqslant  
      }{}
  \fi
\fi

\ifCUPmtlplainloaded \else
  \IfFileExists{amsbsy.sty}
    {\typeout{^^JFound the 'amsbsy' package on the system, using it.^^J}%
     \usepackage{amsbsy}}
    {\providecommand\boldsymbol[1]{\mbox{\boldmath $##1$}}}
\fi

\providecommand\bnabla{\boldsymbol{\nabla}}
\providecommand\bcdot{\boldsymbol{\cdot}}

\newcommand\Real{\mbox{Re}} 
\newcommand\Imag{\mbox{Im}} 

\newsavebox{\astrutbox}
\sbox{\astrutbox}{\rule[-5pt]{0pt}{20pt}}

\newcommand\pd{\ensuremath{\partial}}

\newcommand\etc{etc.\ }
\newcommand\eg{\textit{e.g.}\ }
\newcommand\ie{i.e.\ }

\newcommand{\dd}[2]{\frac{\mathrm{d}#1}{\mathrm{d}#2}}

\newcommand{\grad}{\bnabla}

\newcommand{\cross}{\times}

\newcommand{\Ex}[1]{(\ref{#1})}

\newcommand{\citepos}[1]{\citeauthor{#1}'s (\citeyear{#1})}

\newcommand{\uv}{\boldsymbol{u}}

\newcommand{\xv}{\boldsymbol{x}}
\newcommand{\zhat}{\boldsymbol{\hat z}}

\newcommand{\bv}[1]{\boldsymbol{#1}}

\newcommand{\wa}{\omega_\mathrm{a}}

\newcommand{\freq}{\omega}

\newcommand{\xiv}{\boldsymbol{\xi}}

\newcommand{\trad}{\tau_\mathrm{rad}}
\newcommand{\tdrag}{\tau_\mathrm{fric}}

\newcommand{\Ord}{\mathcal{O}}
\newcommand{\thetat}{\theta}

\newcommand{\LD}{L_\mathrm{D}}
\newcommand{\Leq}{L_\mathrm{eq}}

\title[Super- and sub-rotating equatorial jets: Newtonian cooling versus Rayleigh friction]{Super- and sub-rotating equatorial jets in shallow water models of Jovian atmospheres: Newtonian cooling versus Rayleigh friction} 

\author[E. S. Warneford and P. J. Dellar]{Emma S. Warneford
and Paul J. Dellar\thanks{Email address for correspondence: dellar@maths.ox.ac.uk}}

\affiliation{OCIAM, Mathematical Institute, University of Oxford, Andrew Wiles Building, Radcliffe Observatory Quarter, Woodstock Road, Oxford, OX2 6GG, UK}

\date{27 August 2013, in revised form 22 January 2014, 28 June 2014, 31 March 2015, 30 December 2016}
\begin{document}

\maketitle

\begin{abstract}

  Numerical simulations of the shallow water equations on rotating
  spheres produce mixtures of robust vortices and alternating zonal
  jets, as seen in the atmospheres of the gas giant planets. However,
  simulations that include Rayleigh friction invariably produce a
  sub-rotating (retrograde) equatorial jet for Jovian parameter
  regimes, whilst observations of Jupiter show a super-rotating
  (prograde) equatorial jet that has persisted over several
  decades. Super-rotating equatorial jets have recently been obtained
  in shallow water simulations that include a Newtonian relaxation of
  perturbations to the layer thickness to model radiative cooling to
  space, and in simulations of the thermal shallow water equations
  that include a similar relaxation term in their temperature
  equation. Simulations of global quasigeostrophic forms of these
  different models produce equatorial jets in the same directions as
  the parent models, suggesting that mechanism responsible for setting
  the direction lies within quasigeostrophic theory. We provide such a
  mechanism by calculating the effective force acting on the thickness-weighted
  zonal mean flow  due to the decay of an equatorially trapped Rossby
  wave. Decay due to Newtonian cooling creates an eastward zonal mean
  flow at the equator, consistent with the formation of a
  super-rotating equatorial jet, while decay due to Rayleigh friction
  leads to a westward zonal mean flow at the equator, consistent with
  the formation of a sub-rotating equatorial jet.  In both cases the
  meridionally integrated zonal mean of the absolute zonal momentum is westward, consistent with
  the standard result that Rossby waves carry westward pseudomomentum,
  but this does not preclude the zonal mean flow being eastward on and
  close to the equator.

\end{abstract}




\nocite{AndrewsMcIntyre76a}

\section{Introduction}\label{sec:Intro}

\begin{figure}
\begin{center}
\includegraphics{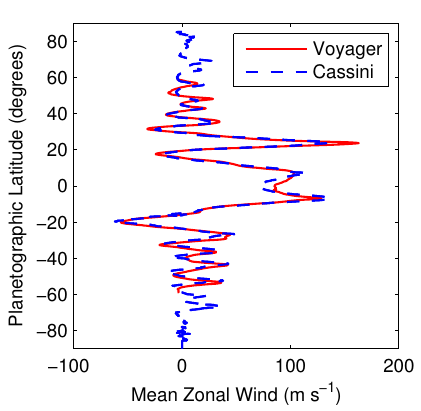} 
\end{center}
\caption{Mean zonal wind profiles for Jupiter. The red solid line shows Voyager 2 data from \cite{limaye1986jupiter} and the blue dashed line shows Cassini data from \cite{porco2003cassini}.}
\label{fig:ubardata}
\end{figure}

Observations of Jupiter's atmosphere reveal a highly turbulent cloud
deck containing long-lived coherent vortices such as the Great Red
Spot that are transported around the planet by an alternating
pattern of  zonal jets \citep{VasavadaShowman05}. 
Following the pioneering non-divergent barotropic model of \cite{Williams78}, shallow water theory has been widely used to model the Jovian atmosphere \citep{Marcus88, DowlingIngersoll89,williams1984geostrophic, ChoPolvani96a, ChoPolvani96b, iacono1999spontaneous, iacono1999high,showman2007numerical, ScottPolvani07, ScottPolvani08, Ingersoll1990}. The visible cloud deck is treated as a homogeneousm dynamically active layer separated from a much deeper and relatively quiescent lower layer by a sharp density contrast, which may be linked to latent heat being released at the water vapour condensation level. The equivalent barotropic approximation \citep{gill82,DowlingIngersoll89} gives a closed set of shallow water equations for the flow in the cloud deck. 

Numerical simulations of the shallow water equations on rotating spheres with Jovian parameter values reproduce a mixture of robust vortices and alternating zonal jets. The latter arise naturally from the nonlinear inverse cascade characteristic of two-dimensional turbulence being arrested through coupling to Rossby waves at the \cite{Rhines75} scale, though the quasilinear ``zonostrophic instability'' mechanism may also have a
role \cite[][and references therein]{SrinivasanYoung12,SrinivasanYoung14}.
The equatorial jets are invariably sub-rotating (retrograde) in both freely decaying and forced-dissipative simulations in the Jovian regime \citep[\eg][]{ChoPolvani96a, ChoPolvani96b, iacono1999spontaneous, iacono1999high, showman2007numerical, ScottPolvani07}. By contrast, the mean zonal wind profiles in figure  \ref{fig:ubardata}, as obtained from
tracking visible features in Jupiter's atmosphere, show a broad, super-rotating equatorial jet. The similarity of the two
profiles, taken from the Voyager 2 \citep{limaye1986jupiter} and Cassini
\citep{porco2003cassini} missions over 20 years apart, demonstrates the
remarkable stability of Jupiter's zonal winds, and the longevity of the super-rotating equatorial jet.

Forced-dissipative shallow water simulations typically include a
Rayleigh friction term to absorb the slow inverse cascade of energy
past the Rhines scale to the largest scales in the system, which tends
to create large coherent vortices in place of the alternating zonal
jets \cite[\eg][]{VasavadaShowman05}. \cite{ScottPolvani07,ScottPolvani08}
added an additional radiative relaxation term with timescale $\trad$ to
the continuity equation in their shallow water model:
\begin{subequations} \label{eq:sp}
\begin{eqnarray}
\label{eq:sp1} h_{t} + \bnabla \bcdot \left(h \bv{u} \right) & = & - \left(h - h_{0}\right) / \trad,\\
\label{eq:sp2} \bv{u}_{t} + \left(\bv{u} \bcdot \bnabla \right) \bv{u} + f \bv{\hat{z}} \times \bv{u} & = & -g' \bnabla h + \bv{F} - \bv{u} / \tdrag,
\end{eqnarray}
\end{subequations}
Here $h$ is the thickness of the active layer, commonly called ``height'', $\bv{u}$ is the depth-averaged horizontal velocity, $f = 2 \Omega \sin \phi$ is the Coriolis parameter at latitude $\phi$ on a planet rotating with angular velocity $\Omega$, $\bnabla$ is the horizontal gradient operator, 
$\bv{\hat{z}}$ is a unit vector in the local vertical direction, $\bv{F}$ is an isotropic random forcing,
and $\tdrag$ is the timescale for Rayleigh friction.
The reduced gravity is $g' = g \Delta \rho / \rho_{0}$, where $g$ is the actual gravitational acceleration, $\Delta \rho$ is the density contrast between the cloud deck and the lower layer, and $\rho_{0}$ is a reference density. These parameters are possibly constrained by  observations of what appear to be internal gravity waves radiating from the impact points of Shoemaker--Levy comet debris \citep{dowling1995estimate, IngersollDowlingGieraschEtAl04}, although this interpretation is not without its critics \citep{Walterscheid2000}.

The right hand side of (\ref{eq:sp1}) models the effects of radiation
to space with a Newtonian relaxation of $h$ towards its constant mean value $h_{0}$ on the time scale $\trad$, following earlier shallow water models of the upper ocean mixed layer \citep{PhilanderYamagataPacanowski84,Hirst86} and the terrestrial stratosphere \citep{Juckes89,PolvaniWaughPlumb95,ThuburnLagneau99}.
Scott \& Polvani's (\citeyear{ScottPolvani07,ScottPolvani08}) simulations of this system with dissipation only by 
thickness relaxation (no $\tdrag$ term) produced a sharply localised super-rotating equatorial jet, as do our subsequent simulations described in \S\ref{sect.num}, and in \cite{WarnefordDellar14tsw}. 

Thermal shallow water theory  introduced by \cite{Lavoie72} to describe atmospheric mixed layers over frozen lakes. It permits horizontal variations of the thermodynamic properties of the fluid within each layer, which are taken to be uniform in the standard shallow water equations. The density contrast $\Delta \rho$ thus becomes a dynamical variable. However, subsequent developments of thermal shallow water theory have been much more focused on modelling
the oceanic mixed layer \citep{cane1983equatorial,Hirst86,mccreary1992equatorial,mccreary1988numerical,mccreary1991numerical,roed1999numerical,Ripa93,Ripa95,Ripa96,Ripa96b}.

\cite{WarnefordDellar14tsw} simulated Jovian atmospheres using the thermal shallow
water equations
\begin{subequations} \label{eq:tsw}
\begin{align} 
\label{eq:tsw1} h_t + \bnabla \bcdot \left(h \bv{u} \right) & = 0,\\
\label{eq:tsw2} \Theta_t + \left( \bv{u} \bcdot \bnabla \right) \Theta & = - \left(\Theta h / h_{0} - \Theta_{0} \right)/\trad,\\
\label{eq:tsw3} {\bv{u}}_t + \left(\bv{u} \bcdot \bnabla \right) \bv{u} + f \boldsymbol{\hat{z}} \times \bv{u} & = - h^{-1}  \bnabla \left( \tfrac{1}{2} \Theta h^2 \right) + \bv{F} - \bv{u} / \tdrag.
\end{align}
\end{subequations}
We introduce the symbol $\Theta = g \Delta \rho / \rho_{0}$ for the reduced gravity to emphasise that it is now a function of space and time. 
The first term on the right hand side of \Ex{eq:tsw3} is then the usual shallow water pressure gradient term, as rewritten for a spatially
varying reduced gravity.
The right hand side of (\ref{eq:tsw2}) is a Newtonian cooling term
that represents radiative relaxation towards a temperature
$\Theta_{0} h_0 / h$ with time scale $\trad$. We show below that this form of coupling between $\Theta$ and $h$ changes the behaviour of equatorially
trapped Rossby waves in almost exactly the same way as the radiative relaxation of $h$
itself in Scott \& Polvani's (\citeyear{ScottPolvani07,ScottPolvani08}) model.
In particular, the damping rates calculated in \S4 are very similar function of wavenumber and $\trad$. The resulting Reynold stresses calculated in \S5 have very similar spatial structures between the two models, and are proportional to the difference between the dimensionless radiative and frictional decay rate constants in both models.

The thermal shallow water equations (\ref{eq:tsw}\textit{a,b,c}) conserve both mass
and momentum in the absence of the $\tdrag$ and  $\bv{F}$ terms, unlike
the Scott and Polvani model  (\ref{eq:sp}\textit{a,b}). Simulations of the thermal
shallow water equations for Jovian parameter values \citep[reported in
\S \ref{sect.num} and][]{WarnefordDellar14tsw} reproduce a
super-rotating equatorial jet, and more substantial mid-latitude jets
than simulations of the \cite{ScottPolvani07,ScottPolvani08}
model. Henceforth, for brevity we refer to the radiative relaxation terms as relaxation and the Rayleigh friction term as friction. We also refer to equations (\ref{eq:sp}\textit{a,b}) as the ``standard'' shallow water equations to distinguish
them from the thermal shallow water equations (\ref{eq:tsw}\textit{a,b,c}).


\label{page:global}

Quasi-geostrophic (QG) theory \citep{Charney49,Obukhov49} offers a simplified description of slow vortical motions that filters out inertia-gravity waves. Although QG theory is usually employed for mid-latitude beta-planes, \cite{Matsuno70,Matsuno71} presented a global QG theory for fully stratified atmospheres on spheres. \cite{verkley09} and \cite{schuberttaftsilvers09} revived this theory for the standard shallow water equations on a sphere in the form
\begin{equation} \label{eq:standard_global_QG}
q_t  + [\psi,q] = 0, \quad q = f + \nabla^2 \psi - \LD^{-2} \psi,
\end{equation}
where the Jacobian $[\psi,q] = \bv{\hat{z}} \cdot ( \bnabla \psi \times \bnabla q)$.
The Rossby deformation scale is $\LD = c/f$, where $c$ is the speed of long surface gravity waves.
Both $f$ and $\LD$ vary with latitude when QG theory is used in spherical coordinates. The sole evolving variable is the potential vorticity $q$,
which is advected by the velocity field derived from a streamfunction $\psi$. The elliptic equation relating $\psi$ to $q$ involves the spatially varying Coriolis parameter $f$. It corresponds to \citepos{Daley83} simplest
form of geostrophic balance, as justified by assuming that $\psi$ varies on lengthscales much smaller than
the planetary scales on which $f$ varies.
The system \Ex{eq:standard_global_QG} reduces to the familiar
beta-plane QG equations on approximating $f$ by $f_0 + \beta y$ in the
first term, and by $f_0$ in the second term.  \cite{ESWthesis} derived
a thermal form of this global QG theory, building on the beta-plane thermal
QG equations in \cite{Ripa96b} and \cite{WarnefordDellarQGTSW}.
Numerical simulations produced super-rotating equatorial jets when the
dimensionless radiative relaxation time $\trad$ is sufficiently short, as did
simulations of the corresponding global QG form of Scott \& Polvani's
(\citeyear{ScottPolvani07,ScottPolvani08}) shallow water
model with a relaxing thickness.  By contrast, both QG models produced sub-rotating equatorial
jets when $\trad$ is sufficiently long. The mechanism responsible for
setting the equatorial jet direction must therefore lie within QG
theory. This focuses attention on momentum fluxes due to Rossby
waves, rather than by inertia-gravity waves, or the equatorially
trapped Kelvin and Yanai waves. None of these waves exist in QG theory.

\section{Momentum transport by Rossby waves}
\label{sect:trans}

\cite{Dickinson69a} showed that the meridional
flux of zonal momentum due to Rossby waves produces a net
eastward acceleration in regions where Rossby waves are generated, and a net westward acceleration in regions
where Rossby waves are dissipated. This argument was further developed by \cite{Green70} and \cite{Thompson71o}.
A detailed version based on ray theory may be found in \cite{Held00whoi} and \cite{vallis2006atmospheric}.
The QG equations on a mid-latitude beta plane lead to the Rossby wave dispersion relation $\omega = -\beta k /(k^2+\ell^2 + \LD^2)$
for plane-wave disturbances with streamfunction $\psi = \hat\psi \exp(\mathrm{i}(k x + \ell y - \omega t))$
 implies a meridional group velocity
\begin{equation}
\frac{\partial \omega}{\partial \ell} = \frac{2 \beta k \ell}{(k^2+\ell^2 + \LD^2)^2},
\end{equation}
while the off-diagonal component of the Reynolds stress (see \S4 for notation) is
\begin{equation}
\langle u' v' \rangle = - \frac{1}{2} \hat \psi^2 k \ell.
\end{equation}
The radiation condition
requires the meridional group velocity to point away from wave sources, so the sign of $k \ell$ is such
as to create a flux of westward (negative) zonal momentum away from sources of Rossby waves towards regions where
Rossby waves are dissipated. 

For example,  \cite{SchneiderLiu09} and \cite{LiuSchneider10} identified Rossby
wave sources due to divergent motions associated with the breakdown of
geostrophy near the equator in their three-dimensional simulations of
Jovian atmospheres. Propagation of these Rossby waves to higher
latitudes thus creates a net eastward acceleration at the equator
according to the above theory, which is consistent with the
formation of super-rotating equatorial jets in their simulations.

However, the focus of our present paper will be on the 
equatorially trapped Rossby waves that play a key
role in near-equatorial dynamics
\cite[\eg][]{Matsuno66,gill82,McCreary85,MajdaKlein03,KhouiderMajdaStechmann13}.
These waves are confined by a Gaussian envelope whose extent scales with the equatorial
deformation scale $\Leq = \sqrt{c/(2\beta)}$. This corresponds to latitudes within about $30^{\circ}$ of the equator for Jovian parameters
(see \S\ref{sec:Rossbywaves}). We therefore need to compute the latitude-dependence
of the momentum fluxes inside the scale of this envelope, for which a ray theory
approach is inadequate.

The generalised Lagrangian mean (GLM) theory is a powerful approach for addressing this
and other problems in wave-mean flow interaction \citep{AndrewsMcIntyre76a,Buhler00,Buhler14book}. It defines
the Lagrangian mean of an arbitrary field $\phi(\xv,t)$ by
\begin{equation}
\phi^\mathrm{L}(\xv,t) = \overline{\phi(\xv+\xiv(\xv,t),t)},
\end{equation}
where $\overline{(\cdots)}$ is some linear averaging operator, and
$\xiv(\xv,t)$ is a Lagrangian displacement field such that
$\xv+\xiv(\xv,t)$ is the actual position of the particle whose mean
position is $\xv$ at time t.  GLM theory establishes circumstances for
the validity of a ``pseudomomentum rule'' \citep{McIntyre81,Buhler00,Buhler14book} under
which the waves behave for some purposes as if they carried a momentum equal to their
pseudomomentum and the background fluid were absent.
The pseudomomentum is a disturbance quantity related to the pseudoenergy
defined in \S5.

A linearized version of GLM for small $\xiv$ was presented in \cite{AndrewsMcIntyre76a, AndrewsMcIntyre78c}. However, the Lagrangian picture of a fluid evolving through displacements of
indestructible particles fits awkwardly with the mass source or
sink term on the right hand side of  \Ex{eq:sp1} that models radiative relaxation. In particular, lack of mass conservation requires modification of B\"uhler's (2000) shallow water pseudomomentum rule obtained from GLM theory. Nevertheless, McIntyre (2014, personal communication) has established, firstly, that is
possible to reproduce our results below using the linearised GLM theory slightly adapted to apply to  (\ref{eq:sp}\textit{a,b})  and, secondly, that an eastward zonal acceleration at the equator is obtained for two interesting cases, (1) equatorially trapped Rossby waves freely decaying by radiative relaxation -- as we compute independently below -- and (2) the same waves, but maintained at constant amplitude by a balance between forcing and dissipation, with the dissipation again by radiative relaxation but with the
forcing mimicked by negative Rayleigh friction. The linearised GLM theory shows \emph{why} negative Rayleigh friction, if spatially uniform, has an effect qualitatively similar to that of free temporal decay -- as illustrated by the curves in figure 4b of \cite{AndrewsMcIntyre76a} with suitable sign changes.


Following the approach of \cite{McIntyreNorton90} for stratified fluids, \cite{Buhler00} formulated a general
approach for shallow water equations of the form
\begin{subequations} \label{eq:MN}
\begin{eqnarray}
\label{eq:MN1} h_{t} + \bnabla \bcdot \left(h \bv{u} \right) & = & 0,\\
\label{eq:MN2} \bv{u}_{t} + \left(\bv{u} \bcdot \bnabla \right) \bv{u} + f \bv{\hat{z}} \times \bv{u} & = & -g' \bnabla h + \bv{F}
\end{eqnarray}
\end{subequations}
with a general term $\mathbf{F}$, not necessarily a forcing. The shallow water potential vorticity $q = (f + \zeta)/h$, where $\zeta = \zhat \cdot (\nabla \cross \uv)$ is the relative vorticity, evolves
according to 
\begin{equation}
\pd_t (h q) + \nabla \cdot( h q \uv - \bv{F}^\perp) = 0,
\end{equation}
with an extra flux $\bv{F}^\perp = \zhat \cross \bv{F}$. \citepos{Buhler00} pseudomomentum rule holds
for systems of this form that conserve linear momentum, those for which $h \bv{F} = \nabla \cdot \boldsymbol{\sigma}$ is the divergence of a stress tensor $\boldsymbol{\sigma}$

However, neither of of the shallow water models from \S1 fits this form. 
The shallow water system with relaxation of the thickness does not conserve momentum. In the absence of forcing and friction,
(\ref{eq:sp1}) and (\ref{eq:sp2}) together imply
\begin{equation}
\left(h\bv{u}\right)_{t} + \nabla \cdot \left(h \bv{u} \bv{u} + \tfrac{1}{2} g' h^2 \bv{I} \right) + f \hat{\bv{z}} \times \left(h\bv{u} \right)  = - \left(h - h_{0} \right) \bv{u}  / \trad, \label{SPmomentum}
\end{equation}
where $\bv{I}$ is the $2 \times 2$ identity matrix.
Moreoever, the potential vorticity for the system (\ref{eq:sp}\textit{a,b}) obeys
\begin{equation}
 (h q)_t + \nabla \cdot ( h q \uv + \uv^\perp/\tdrag ) = 0, 
\end{equation}
where $\uv^\perp =  \zhat \cross \uv$, 
while the potential vorticity in the 
thermal shallow water equations (\ref{eq:tsw}\textit{a--c}) obeys
\begin{equation}
(h q)_t + \nabla \cdot( h q \uv - \tfrac{1}{2} h \grad^\perp \theta + \uv^\perp/\tdrag ) = 0. \label{TSWpv}
\end{equation}
Radiative relaxation thus does not contribute to either potential
vorticity equation. However, \Ex{TSWpv} contains an extra term
involving the perpendicular temperature gradient $\grad^\perp \theta =
\zhat \cross \grad \theta$ that arises from the curl of the $(1/2) h
\grad \Theta$ term on the right hand side of \Ex{eq:tsw3}. The thermal
shallow water equations thus do not materially conserve potential
vorticity, even in the absence of forcing and dissipation. Instead,
the total potential vorticity inside a closed isotherm is conserved.
This is the thermal shallow water analogue of the potential vorticity impermeability property of
isentropic surfaces in three-dimensional stratified fluids
\citep{HaynesMcIntyre90}.

Since the \cite{ScottPolvani07,ScottPolvani08} radiatively relaxed shallow water model lacks both mass and momentum
conservation, and hence does not fit naturally into the above general frameworks, in \S\ref{sec:Rossbywaves} and \S\ref{sect:mom}
below we derive from first principles equations for the evolution of the thickness-weighted zonal mean velocity components $\langle u \rangle^*$
and  $\langle v \rangle^*$ (see \S5 for notation) caused by the decay of an
equatorially trapped Rossby wave by friction and/or radiative relaxation.

\section{Numerical experiments} 
\label{sect.num}

\begin{table}
  \begin{center}
\def~{\hphantom{0}}
  \begin{tabular}{cccccc}
  		$a$ \quad \quad & $2 \pi / \Omega$ \quad \quad & $\sqrt{g' h_{0}}$ \quad \quad & $g$ \quad \quad & $\trad$ 
  			\quad \quad & $\tdrag$ \\
		$7.1  \times 10^7$m \quad \quad & 9.9 hours \quad \quad & 678 $\mathrm{m \, s}^{-1}$ \quad \quad & 26 $\mathrm{m \, s}^{-2}$ \quad \quad & 429 hours \quad \quad & 9900 hours \\
  \end{tabular}
  \caption{Parameter values for Jupiter \citep[from][]{Ingersoll1990,Beebe1994,WarnefordDellar14tsw}.}
  \label{tab:Jupdata}
  \end{center}
\end{table}

We now show some numerical simulations of the standard shallow water
equations with dissipation only through friction (\ie
(\ref{eq:sp}\textit{a,b}) with no $\trad$ term), the
\cite{ScottPolvani07,ScottPolvani08} model with relaxation of the
thickness, and our thermal shallow water model. Table
\ref{tab:Jupdata} gives typical Jovian values for the planetary radius
$a$, angular velocity $\Omega$, gravitational acceleration $g$, and
internal wave speed $\sqrt{g' h_{0}}$. The last is deduced from
impacts of Shoemaker--Levy comet debris \citep{dowling1995estimate,
  IngersollDowlingGieraschEtAl04}, and implies that there are $232$
polar deformation scales $\LD=\sqrt{g' h_0}/(2 \Omega)$ around the
circumference. Simulations with radiative relaxation used a timescale
$\trad$ of 43 Jovian days, as calculated for a temperature of $120$K
at the 25 millibar pressure level, and all simulations employed a
friction with a time scale of 1000 Jovian days. \cite{LiuSchneider10}
used the 25 millibar pressure level when comparing their
three-dimensional simulations with observations of
Neptune. Calculating $\trad$ at the 25 millibar pressure level for
both planets leads to a super-rotating equatorial jet for Jupiter, and
a sub-rotating equatorial jet for Neptune, in our thermal shallow
water model. The theory in \S5 below suggests that the direction of
the equatorial jet is insensitive to these parameters provived the
radiative relaxation timescale remains much shorter than the
frictional timescale.

\begin{figure}
\begin{center}
\includegraphics{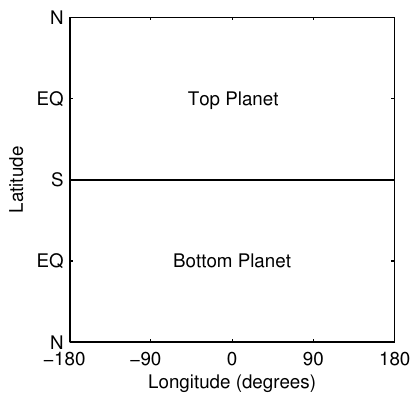}
\end{center}
\caption{Schematic of the doubly-periodic Cartesian domain used for the numerical simulations. Each half of the domain represents a full planet, so for ease of interpretation we show simulation results for just the top planet.}
\label{fig:square}
\end{figure}

Simulations of geostrophic turbulence in the Jovian regime require thousands of rotation periods to reach statistically steady states. The superior memory bandwidth and peak floating point performance of graphical processing units (GPUs) thus becomes attractive, but existing spherical harmonic transform algorithms for GPUs only achieve performance parity with conventional microprocessors \citep{HupcaFalcouGrigoriStompor12}. We therefore employed the doubly-periodic Cartesian domain sketched in figure \ref{fig:square}, which we refer to as the square planet domain. The horizontal axis denotes longitude, while the vertical axis denotes latitude. Starting from the north pole at the top of the domain and moving down, we reach the equator a quarter of the way down, and the south pole half way down. We then continue to reach the equator again, before returning to the north pole at the bottom of the domain.  We imagine following a meridian (constant longitude line) from the north pole to the equator to the south pole, and then following a second meridian with longitude offset by $180^{\circ}$ back across the equator to the north pole. The Coriolis force is still fully varying in latitude and is defined in the simulations by $f(y) = 2 \Omega \cos(2 \pi y / y_\mathrm{max})$, where $y_\mathrm{max}$ is the length of the side of the square planet domain, \ie the circumference of Jupiter (see table \ref{tab:Jupdata}). The Coriolis force and all its derivatives are thus doubly periodic. This allowed us to exploit a high performance GPU fast Fourier transform library. We discretized the domain
using $1024 \times 1024$ Fourier collocation points, and used the \cite{hou2007computing} spectral filter to control the build-up of enstrophy at the highest wavenumbers. We  integrated the resulting large system of ordinary differential equations using the standard fourth-order Runge--Kutta scheme, with a time step determined dynamically from the Courant--Friedrichs--Lewy stability condition.

Our initial conditions were $h = h_{0}$ and $u = v = 0$ for all runs, and $\Theta=\Theta_{0}$ for the thermal shallow water simulation.
Following  \cite{ScottPolvani07,ScottPolvani08} we applied a divergence-free isotropic random forcing $\bv{F}$ that is localised to a narrow  annulus of wavenumbers $|\mathbf{k}| \in [40,44]$ in Fourier space, counting the longest sinusoidal mode in the domain as wavenumber $1$.
We forced each mode inside this annulus with amplitude $\epsilon_{f}$ using random phases that were $\delta$-correlated (white) in time. 
This forcing is widely used in numerical studies of zonal jet formation, and models the energy injected by three-dimensional convection at horizontal lengthscales comparable to the deformation radius. Our simulations slowly adjusted the amplitude of the forcing $\epsilon_{f}$ to reach a prescribed value of the total kinetic energy in the eventual statistical steady state. Further details of the numerical models, parameters, and simulation outputs may be found in  \cite{ESWthesis} and \cite{WarnefordDellar14tsw}.

\begin{figure}
\begin{center}
   (a)\includegraphics[bb=110 270 522 450,clip]{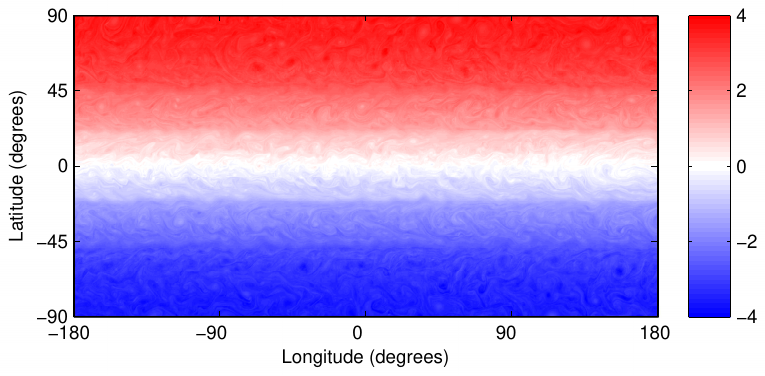}
   (b)\includegraphics[bb=110 270 522 450,clip]{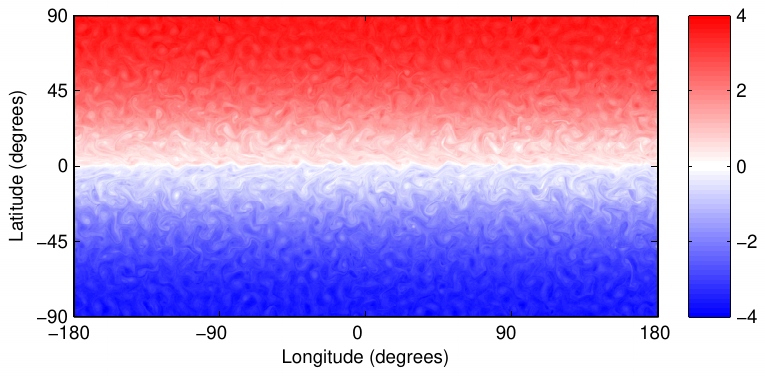}
   (c)\includegraphics[bb=110 270 522 450,clip]{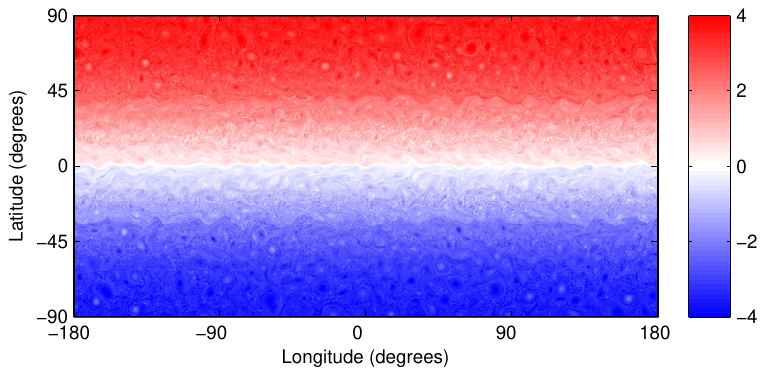}
\end{center}
\caption{Absolute vorticity $\wa = v_x - u_y + f(y)$ in units of $10^{-4}$ s$^{-1}$ for (a) the standard shallow water equations with dissipation by friction alone, (b) 
the standard shallow water equations with friction and radiative relaxation of the thickness, and  (c) the thermal shallow water equations with friction and radiative relaxation of the temperature.}
\label{fig:Z}
\end{figure}

\begin{figure}
\begin{center}
   (a)\includegraphics[bb=110 270 522 450,clip]{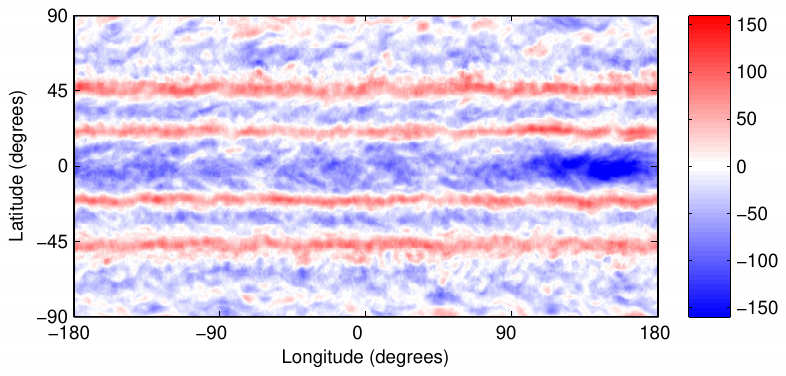}
   (b)\includegraphics[bb=110 270 522 450,clip]{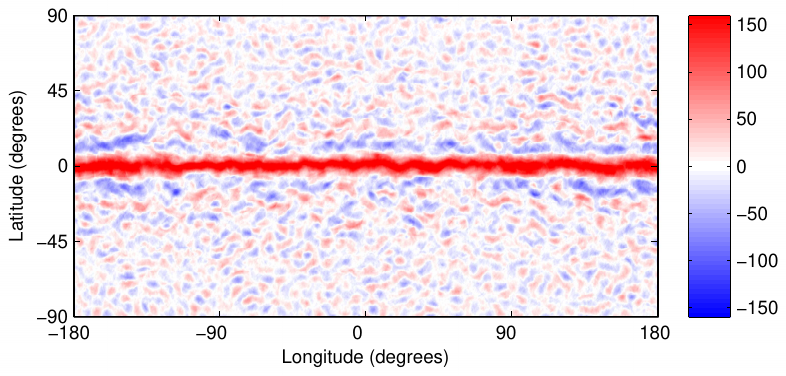}
   (c)\includegraphics[bb=110 270 522 450,clip]{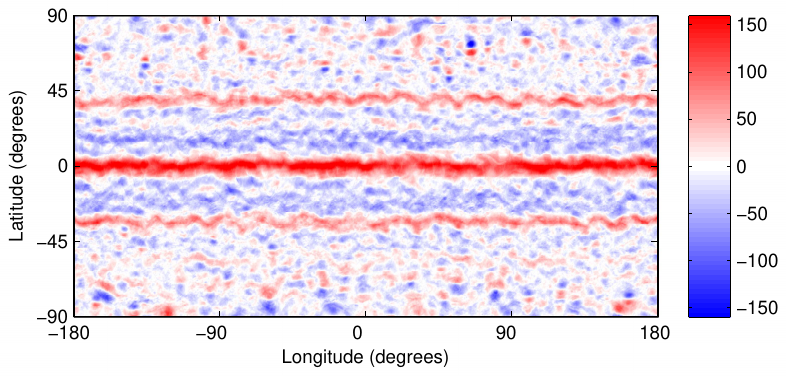}
\end{center}
\caption{Zonal velocity $u$ in units of m s$^{-1}$ for (a) the standard shallow water equations with dissipation by friction alone, (b) 
the standard shallow water equations with friction and radiative relaxation of the thickness, and  (c) the thermal shallow water equations with friction and radiative relaxation of the temperature.}
\label{fig:U}
\end{figure}

\begin{figure}
\begin{center}
\includegraphics{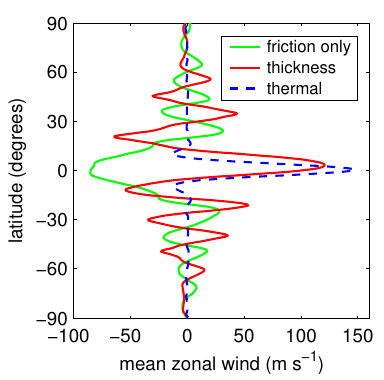} 
\end{center}
\caption{Mean zonal velocity $\langle u \rangle$ against latitude for simulations of the standard shallow water equations with dissipation by friction alone, the standard shallow water equations with friction and radiative relaxation of the thickness, and the thermal shallow water equations with friction and radiative relaxation of the temperature}
\label{fig:ubar}
\end{figure}

Figure \ref{fig:Z} shows the instantaneous absolute vorticity $\wa = v_{x} - u_{y} + f(y)$ after $2 \times 10^4$ Jovian rotation periods for the top planet in figure \ref{fig:square} for three different simulations: the standard shallow water equations with friction alone, the standard shallow water equations with friction and radiative relaxation of the thickness, and the thermal shallow water equations with friction and radiative relaxation of the temperature.
Figure \ref{fig:U} shows the corresponding instantaneous zonal velocity plots, and figure \ref{fig:ubar} shows the  instantaneous zonally averaged zonal velocity, $\langle u \rangle  = y_\mathrm{max}^{-1} \int u(x, y, t) \, \mbox{d}x$. All three simulation runs exhibit a mixture of vortices, turbulence and multiple zonal jets, with amplitudes that decrease at higher latitudes. Both simulations with radiative relaxation produced a strong super-rotating equatorial jet in line with observations of Jupiter. However, our simulation of the standard shallow water model, with dissipation only by friction, produced a sub-rotating equatorial jet. The simulation with radiative relaxation of the thickness produced very weak jets away from the equator, whereas the other two simulations produced stronger mid-latitude jets in better agreement with observations.

\section{Rossby waves on an equatorial beta-plane} \label{sec:Rossbywaves}

We now investigate the properties of equatorially trapped waves on an
equatorial beta plane in the different shallow water models, as they
decay freely due to Rayleigh friction and/or Newtonian radiative
relaxation. The equatorial beta plane uses local Cartesian coordinates
with $x$ eastward and $y$ northward. The underlying spherical geometry
appears only through the linearisation of $f = 2 \Omega \sin \phi$ for
small latitudes $|\phi| \ll1$ into $f(y) = \beta y$ with $\beta = 2
\Omega / a$, where $a$ is the planetary radius.

\subsection{Shallow water equations with thickness relaxation} \label{subsec:Rossbywaves_h}
The linearised form of the unforced ($\bv{F}=\bv{0}$) shallow water equations \Ex{eq:sp} with radiative relaxation of the thickness on an equatorial beta-plane may be written as \citep{Matsuno66, gill82}
\begin{subequations}
\label{LinBetaNonDim}
\begin{eqnarray}
\label{eq:l3} h'_t + u'_x + v'_y &=& - \kappa h', \\
\label{eq:l1} u'_t - \tfrac{1}{2} y v' + h'_x &=& - \gamma u', \\
\label{eq:l2} v'_t + \tfrac{1}{2} y u' +  h'_y &=& - \gamma v'.
\end{eqnarray}
\end{subequations}
The dashes denote small perturbations about a rest state with uniform
depth $h_{0}$.  We have non-dimensionalised using the internal wave
speed $c=\sqrt{g'h_{0}}$ as the velocity scale, the equatorial
deformation radius $L_\mathrm{eq} = \sqrt{c/2\beta}$ as the horizontal
length scale, and $h_{0}$ as the thickness scale. Using the Jovian
parameters in table~\ref{tab:Jupdata}, a latitude of $33^\circ$ corresponds
to $y=5$ in the subsequent plots.
The dimensionless
radiative relaxation and friction rates are $\kappa = T / \trad$ and $\gamma = T /
\tdrag$ based on the advective time scale $T = L_\mathrm{eq} /
c$. Their values were  $\kappa = 7.88 \times 10^{-3}$ and $\gamma = T / \tdrag = 3.42 \times 10^{-4}$ in the simulation with
radiative relaxation of the thickness.

\begin{figure}
\begin{center}
\includegraphics{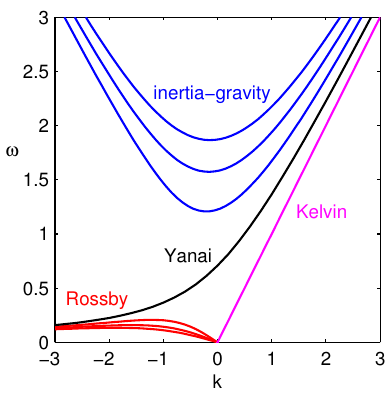}
\end{center}
\caption{Dimensionless dispersion relation for equatorially trapped waves in the standard
shallow water equations on the equatorial beta-plane with no dissipation.
}
\label{fig:eqwaves}
\end{figure}

Following standard practice \cite[\eg\ ][]{Matsuno66,gill82} we seek waves that are harmonic in longitude and time of the form 
$h'(x,y,t)= \Real \{ \hat h(y)\exp(\mathrm{i} (kx-\freq t)) \}$ \etc
With no  dissipation  ($\kappa = \gamma = 0$) 
the three equations (\ref{LinBetaNonDim}) may be combined into a single ordinary differential
equation (ODE) for the meridional velocity $\hat v(y)$,
\begin{equation}
\dd{^2 \hat v}{y^2} = (A y^2 - B) \hat v, \quad
A = \frac{1}{4}, \quad
B = \freq^2 - k^2 - \frac{k}{2 \freq},
\label{TrapODE}
\end{equation}
the same equation that governs a quantum harmonic oscillator. The solutions that decay as $y \to \pm \infty$ may be written using the Hermite polynomials $H_n(\xi)$ as 
\begin{equation}
\hat v  = H_n ( \xi ) \exp(- \xi^2/2), \quad \xi = y A^{1/4}.
\label{TrapHermite}
\end{equation}
The dispersion relation $A^{-1/2} B = 2 n+1$ for $n=-1,0,1,2,\ldots$ gives a cubic
equation for $\freq$,
\begin{equation}\label{eq:dr1}
\freq^2 - k^2 - \frac{k}{2\freq} 
=  n+\frac{1}{2}.
\end{equation}
The Rossby wave branch of solutions is characterised by $n \ge 1$ and $0 < - \freq/ k \ll 1$. The other two roots
give inertia-gravity waves with $|\freq| \ge 1$ (the dimensionless inertial frequency). The cases $n=0$ and $n=-1$ give the
Yanai and Kelvin waves respectively.  Figure~\ref{fig:eqwaves} shows these different branches
of the dispersion relation, all of which represent trapped waves localised within a few equatorial deformation
scales $\Leq$ of the equator. The  inertia-gravity, Yanai, and Kelvin wave branches all exceed the inertial
frequency $|\omega| =1$, and so do not exist in QG theory. 
The mechanism responsible for setting the direction of equatorial jet in the shallow water models is expected to lie within QG theory, so we only consider the Rossby waves in the remainder of this paper.

The friction and radiative relaxation terms in (\ref{LinBetaNonDim}\textit{a--c})
change the $A$ and $B$ coefficients to \citep{YamagataPhilander85}
\begin{equation} \label{eq:AB}
A = \frac{1}{4} \left( \frac{\freq + \mathrm{i} \kappa}{\freq + \mathrm{i} \gamma} \right), \quad B = \freq^2 - k^2  - \frac{k}{2\left(\freq + \mathrm{i} \gamma\right)} + \mathrm{i} \freq \left(\kappa + \gamma \right) - \kappa \gamma,
\end{equation}
so the dispersion relation becomes
\begin{equation}\label{eq:dr2}
\freq^2 - k^2 - \frac{k}{2(\freq + \mathrm{i} \gamma)} 
+ \mathrm{i} \freq (\kappa + \gamma) - \kappa \gamma
=  \left(  n+\frac{1}{2} \right) \left( \frac{\freq + \mathrm{i} \kappa}{\freq + \mathrm{i} \gamma} \right)^{1/2} \mbox{ for } n=1,2,\ldots.
\end{equation}

\begin{figure}
\begin{center}
(a)\includegraphics{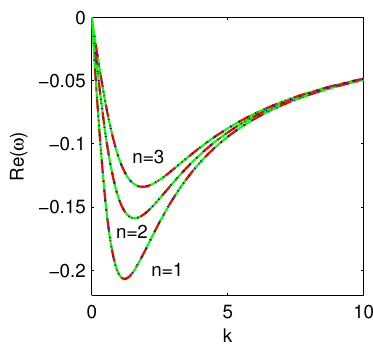}
(b)\includegraphics{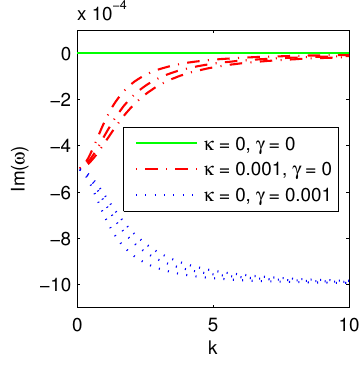}
\end{center}
\caption{The (a) real and (b) imaginary parts of the dimensionless equatorial Rossby wave frequency $\freq$ for the standard shallow water equations with no dissipation (green solid lines), dissipation by relaxation of the thickness (red dash-dot lines), and dissipation by friction (blue dotted lines).}
\label{fig:dr}
\end{figure}

Figure \ref{fig:dr} shows the real and imaginary parts of the complex
frequency $\freq$ for the first few equatorial Rossby waves
($n=1,2,3$) for different values of $\gamma$ and $\kappa$. The
dispersion relation (\ref{eq:dr1}) is invariant under the transformation $(k,\omega) \mapsto (-k,-\omega)$,
so we choose to take $k>0$. The Rossby wave frequency $\Real(\omega)$ is now negative, and virtually
unchanged by either weak friction ($\kappa=0.001$, $\gamma=0$) or weak radiative relaxation ($\kappa=0$, $\gamma=0.001$). The decay rate $- \Imag(\omega)$ due to cooling ($\kappa=0.001$, $\gamma=0$) is largest at $k=0$, and tends to zero as $k \to \infty$. By contrast, the decay due to friction ($\gamma=0.001$, $\kappa=0$) is monotonically increasing with $k$, and $-\Imag(\omega)$ tends to $-\gamma$ as $k \to \infty$. 
Figure \ref{fig:dr}b also suggests a symmetry about the line $\Imag (\freq) = - \epsilon / 2$ between small radiative relaxation ($\kappa=\epsilon$ and $\gamma=0$) and small friction ($\kappa=0$ and $\gamma=\epsilon$). We expand the roots of (\ref{eq:dr2}) as \label{pert} $\freq = \freq_{0} + \epsilon \, \freq_{1,\mathrm{rad}} + \epsilon^2 \, \freq_{2,\mathrm{rad}} + \cdots$ for radiative relaxation, and $\freq = \freq_{0} + \epsilon \, \freq_{1,\mathrm{fric}} + \epsilon^2 \, \freq_{2,\mathrm{fric}} + \cdots$ for friction, where  $\freq_0$ is the real root of \Ex{eq:dr1}. The $O(\epsilon)$ corrections are both purely imaginary:
\begin{subequations}
\begin{eqnarray}
\freq_{1,\mathrm{rad}} & = & \tfrac{1}{2} \mathrm{i} \freq_{0} ( - 4 \freq_{0}^2 + 2n + 1)/(k + 4 \freq_{0}^3 ),\\
\freq_{1, \mathrm{fric}} & = & \tfrac{1}{2} \mathrm{i} \freq_{0} ( - 4 \freq_{0}^2 - 2n - 1 - 2 k / \freq_{0}) / (k + 4 \freq_{0}^3 ).
\end{eqnarray}
\end{subequations}
The symmetry about $\Imag (\freq) = - \epsilon / 2$ visible in figure \ref{fig:dr}b is a consequence of
the relation
\begin{equation}
\freq_{1,\mathrm{rad}} + \frac{\mathrm{i}}{2} = \frac{\mathrm{i}}{2} \left( \frac{k + (2n+1) \omega_{0}}{k + 4 \omega_{0}^3 }\right) = - \left( \freq_{1,\mathrm{fric}} + \frac{\mathrm{i}}{2} \right)
\end{equation}
between the two dispersion relations truncated at $O(\epsilon)$.


\subsection{Thermal shallow water equations} \label{subsec:Rossbywaves_t}

\begin{figure}
\begin{center}
(a)\includegraphics{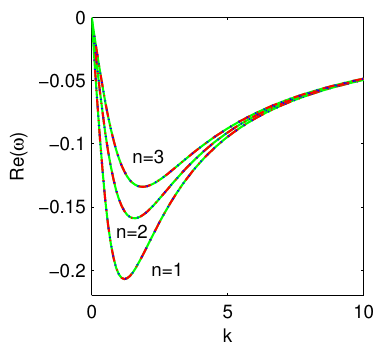}
(b)\includegraphics{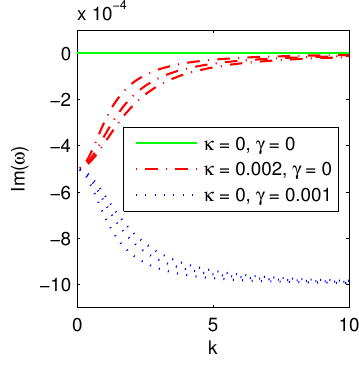}
\end{center}
\caption{The real and imaginary parts of the dimensionless equatorial Rossby wave frequency $\freq$ for the thermal shallow water equations with no dissipation (green solid lines), dissipation by relaxation of the temperature (red dash-dot lines), and dissipation by friction (blue dotted lines).}
\label{fig:drT}
\end{figure}

Linearising the thermal shallow water equations (\ref{eq:tsw}\textit{a--c}) about a rest state with uniform depth $h_{0}$ and temperature $\Theta_{0}$ and applying the equivalent nondimensionalisation based on the
velocity scale $\sqrt{h_0 \Theta_0}$ gives
\begin{subequations}
\label{LinBetaNonDimT}
\begin{eqnarray}
\label{eq:Tl3} h'_t + u'_x + v'_y &=& 0, \\
\label{eq:Tl4} \Theta'_{t} &=& - \kappa \left(h' + \Theta'\right),\\
\label{eq:Tl1} u'_t - \tfrac{1}{2} y v' + h'_x + \tfrac{1}{2} \Theta'_{x} &=& - \gamma u' , \\
\label{eq:Tl2} v'_t + \tfrac{1}{2} y u' +  h'_y + \tfrac{1}{2} \Theta'_{y} &=& - \gamma v' .
\end{eqnarray}
\end{subequations}
The four equations (\ref{LinBetaNonDimT}) may again be combined into
the single ODE \Ex{TrapODE} for $\hat{v}(y)$, with coefficients
\begin{equation}
\label{eq:AB_thermal}
A = \frac{1}{4} \frac{\freq \left( \freq + \mathrm{i} \kappa \right)}{\left( \freq + \mathrm{i} \gamma \right) \left( \freq + \mathrm{i} \kappa / 2 \right)}, \quad B = \frac{\freq \left(\freq + \mathrm{i} \gamma \right) \left(\freq + \mathrm{i} \kappa \right)}{\freq + \mathrm{i} \kappa / 2} - k^2 - \frac{k}{2 \left( \freq + \mathrm{i} \gamma \right)},
\end{equation}
so the dispersion relation for equatorially trapped waves is
\begin{equation}
\label{eq:drT}
\frac{\freq \left(\freq + \mathrm{i} \gamma \right) \left(\freq + \mathrm{i} \kappa \right)}{\freq + \mathrm{i} \kappa / 2} - k^2 - \frac{k}{2 \left( \freq + \mathrm{i} \gamma \right)} = \left(n + \frac{1}{2}\right) \left( \frac{\freq \left(\freq + \mathrm{i}\kappa\right)}{\left(\freq + \mathrm{i} \gamma\right)\left(\freq + \mathrm{i} \kappa / 2 \right) } \right)^{1/2}.
\end{equation}
Figure \ref{fig:drT} shows the real and imaginary parts of the complex
frequency $\freq$ for the first few equatorial Rossby waves with $n \in \{1,2,3\}$,
and three different sets of $\kappa$ and $\gamma$
values. As before, the real part of $\freq$ is virtually unchanged by weak
dissipation ($\kappa=0.002$ or $\gamma=0.001$) while $\Imag (\freq )$
becomes negative. The plot of $\Imag (\freq )$ in figure \ref{fig:drT} closely resembles our earlier
figure \ref{fig:dr}, with the same apparent symmetry, if we now take $\kappa = 2 \gamma$. A
similar perturbation analysis to before shows that including this factor of $2$ indeed makes $\Imag(\freq)$
symmetric about $-\epsilon / 2$ between the cases of small radiative relaxation
($\kappa = 2 \epsilon$ and $\gamma = 0$) and small friction ($\kappa =
0$ and $\gamma = \epsilon$).

\section{Acceleration of the zonal mean zonal flow} \label{sect:mom}
We now investigate the acceleration of the zonal mean zonal flow caused by these
solutions for decaying equatorially trapped Rossby waves.

\subsection{Shallow water equations with thickness relaxation} \label{subsec:mom_h}
We consider the shallow water model (\ref{eq:sp}) with radiative relaxation of the thickness, and no forcing, on an equatorial beta-plane, and apply
the nondimensionalisation from \S\ref{subsec:Rossbywaves_h} to obtain
\begin{align}
h_t + (hu)_x + (hv)_y &= - \kappa ( h - 1), \label{eq:hflux} \\
\left( h u \right)_{t} + \left(h u^2 + \tfrac{1}{2} h^{2} \right)_{x} + \left( h u v \right)_{y} - \tfrac{1}{2} y h v &= -\gamma h u - \kappa u \left(h - 1 \right), \label{eq:uflux} \\
\left( h v \right)_{t} + \left( h u v \right)_{x} + \left(h v^2 + \tfrac{1}{2} h^{2} \right)_{y} + \tfrac{1}{2} y h u &= -\gamma h v - \kappa v \left(h - 1 \right). \label{eq:vflux}
\end{align}
The mean thickness $h_0$ becomes unity under this nondimensionalisation.

Following standard practice, we decompose the velocity and other fields as $\bv{u} = \langle \bv{u} \rangle + \bv{u}'$ into a zonal mean $\langle \bv{u} \rangle$ and a deviation $\bv{u}'$. The zonal mean of the dimensionless mass conservation equation may then be written as
\begin{equation} \label{eq:nondimmass}
\langle h \rangle_{t} + \langle h v \rangle_{y} = -\kappa \left(\langle h \rangle - 1 \right).
\end{equation}
Similarly, the zonal means of the two momentum equations are
\begin{equation}
\langle h u \rangle_{t} + \langle h u v \rangle_{y} - \tfrac{1}{2} y \langle h v \rangle = -\gamma \langle h u \rangle - \kappa \left( \langle h u \rangle - \langle u \rangle \right), \label{eq:umflux} \\
\end{equation}
and
\begin{equation}
\langle h v \rangle_{t} + \langle h v^2 + \tfrac{1}{2} h^{2}  \rangle_{y} + \tfrac{1}{2} y \langle h u \rangle = -\gamma \langle h v \rangle - \kappa \left( \langle h v \rangle - \langle v \rangle \right). \label{eq:vmflux} \\
\end{equation}

We now consider the form of solutions to the system \Ex{eq:hflux} to \Ex{eq:vflux} for initial conditions corresponding to a Rossby wave of small
amplitude $\Ord(a)$ superimposed on a rest state with unit thickness. For these initial conditions, and under the subsequent evolution, the thickness and velocity components satisfy the scalings
\begin{equation}
h = 1 + \underbrace{h'}_{\Ord(a)} + \underbrace{\left( \langle h \rangle - 1 \right)}_{\Ord(a^2)}, \quad
u = \underbrace{u'}_{\Ord(a)} + \underbrace{\langle u \rangle}_{\Ord(a^2)}, \quad
v = \underbrace{v'}_{\Ord(a)} + \underbrace{\langle v \rangle}_{\Ord(a^2)}. \label{Scalings}
\end{equation}
We define  $\langle \eta \rangle^{*} = \langle h \eta \rangle / \langle h \rangle$ to be the thickness-weighted zonal mean of a quantity $\eta$, as in \cite{ThuburnLagneau99}. Using $\langle h \rangle = 1 + \Ord(a^2)$, we can simplify the above to obtain evolution equations
correct to $\Ord(a^2)$ in the form
\begin{align}
\langle h \rangle_t + \langle v \rangle^{*}_{y} + \kappa \left(\langle h \rangle - 1 \right) &= 0, \label{Hmean} \\
\langle u \rangle^{*}_t - \tfrac{1}{2} y \langle v \rangle^{*} + \gamma \langle u \rangle^{*}
&= - \langle u' v' \rangle_y - \kappa \langle h' u' \rangle , \label{Umean} \\
\langle v \rangle^{*}_t + \tfrac{1}{2} y \langle u \rangle^{*} + \gamma \langle v \rangle^{*} + \langle h \rangle_y
&=   - \langle v' v' \rangle_y - \kappa \langle h' v' \rangle - \tfrac{1}{2} \langle h'^2 \rangle_y . \label{Vmean}
\end{align}
On the left hand sides we have only the three mean quantities $\langle
h \rangle$, $\langle u \rangle^{*}$, $\langle v \rangle^{*}$.  On the
right hand sides we have only means of quadratic products of the fluctuations
$h'$, $u'$, $v'$. These include the meridional derivatives of the Reynolds stress components $ \langle u' v' \rangle$ and $\langle v' v' \rangle$.
The contribution $- \tfrac{1}{2} \langle h'^2 \rangle_y $ from the fluctuating pressure gradient arises from using \Ex{Scalings}
to write
\begin{equation}
h^2 = 1 + 2 h' + 2 \left( \langle h \rangle - 1 \right) + h'^2 + \Ord(a^3).
\end{equation}
There are also radiative relaxation terms proportional to the so-called bolus velocity components $\langle h' u' \rangle$
and $\langle h' v' \rangle$. These terms would be absent for the momentum-conserving variant of the
shallow water equations with relaxation of the thickness.

The effect of the Rossby wave thus appears as an
effective force on the right hand sides of the evolution equations for $\langle u \rangle^{*}$
and $\langle v \rangle^{*}$. Our use of the thickness-weighted means $\langle u \rangle^{*}$
and $\langle v \rangle^{*}$ in place of the unweighted means $\langle u \rangle$ and $\langle v \rangle$
eliminates a $\langle h' v' \rangle$ term that would otherwise appear in \Ex{Hmean}. This is similar
to the transformed Eulerian mean (TEM) approach for stratified fluids \citep{AndrewsMcIntyre76a,AndrewsMcIntyre78c}.

 A further simplification is possible by assuming that
geostrophic balance holds in the meridional momentum equation \Ex{Vmean}, since the meridional lengthscale
and velocity component are both smaller than the zonal lengthscale and zonal velocity component. The same scaling
argument is used in the semigeostrophic theory of atmospheric front formation.
The time derivative of this meridional geostrophic balance condition gives
\begin{equation}
\tfrac{1}{2} y \langle u \rangle^{*}_t + \langle h \rangle_{yt} =  - \tfrac{1}{2} \langle h'^2 \rangle_{yt}, \label{VmeanG}
\end{equation}
which can be used in place of \Ex{Vmean} to obtain a closed system with \Ex{Hmean} and \Ex{Umean}.

We now evaluate these right hand sides using the equatorially trapped Rossby wave solutions
in \S \ref{subsec:Rossbywaves_h}. These represent small perturbations about a state of rest with unit depth in dimensionless variables. The meridional velocity perturbation is $v' = \alpha \, \Real \{ \hat{v}(y) \exp(\mathrm{i} (k x-\freq t)) \}$, with $\hat{v}(y)$  defined by \Ex{TrapHermite} and (\ref{eq:AB}), and the constant $\alpha$ sets the amplitude. The corresponding zonal velocity and thickness perturbations are $u' = \alpha \, \Real \{ \hat{u}(y) \exp(\mathrm{i} (k x-\freq t)) \}$ and $h' = \alpha \, \Real \{ \hat{h}(y) \exp(\mathrm{i} (k x-\freq t)) \}$, where
\begin{equation} \label{eq:uhat}
\hat{u}(y) = \mathrm{i} \left( \frac{ k \left(\mbox{d}\hat{v} / \mbox{d} y\right) - \tfrac{1}{2} y \left( \freq + \mathrm{i} \kappa \right) \hat{v}}{k^2 - \left( \freq + \mathrm{i} \gamma \right) \left( \freq + \mathrm{i} \kappa \right)} \right), \quad
\hat{h}(y) = - \mathrm{i} \left( \frac{\left( \mathrm{i} \freq  - \gamma \right) \hat{u}  + \tfrac{1}{2} y \hat{v}}{k} \right),
\end{equation}
for $k \neq 0$. The denominator in the expression for  $\hat{u}(y)$ vanishes when $k^2 = \left( \freq + \mathrm{i} \gamma \right) \left( \freq + \mathrm{i} \kappa \right)$. This gives the dispersion relation for equatorially trapped Kelvin waves, which are distinguished by having zero meridional velocity ($\hat{v} = 0$), and frequencies $\omega = \pm k$ in the absence of dissipation.

To allow a fair comparison between the effective forces generated by waves with different $n$ and $k$ values, we normalise the
disturbance amplitudes by choosing $\alpha$ so that
\begin{equation}
\frac{1}{2} \int \langle u'^2 \rangle + \langle v'^2 \rangle + \langle h'^2 \rangle \, \mbox{d} y = a^2. \label{eq:PseudoSW}
\end{equation}
The left hand side of \Ex{eq:PseudoSW} is the quadratic approximation
to the disturbance pseudoenergy, an exact conserved quantity of the
unforced, non-dissipative shallow water equations that takes the above
form when expanded for small-amplitude disturbances from a rest state
of constant depth \citep{Shepherd93}. Using a standard formula for
the average of a product of sinusoidal disturbances represented by the
real parts of complex exponentials, we can compute the integrand in
\Ex{eq:PseudoSW}, and the terms on the right hand sides of \Ex{Umean}
and \Ex{Vmean}, using
\begin{equation}
\langle u' v' \rangle = \tfrac{1}{2} \, \Real \left( \hat{u}(y) \hat{v}(y)^\dagger \right)
\end{equation}
where the superscript ${}^\dagger$ denotes a complex conjugate, and similarly for the other quadratic combinations of $h'$, $u'$, $v'$.

\begin{figure}
\begin{center}
(a) \includegraphics{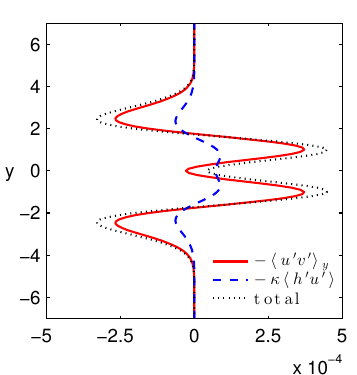} 
(b) \includegraphics{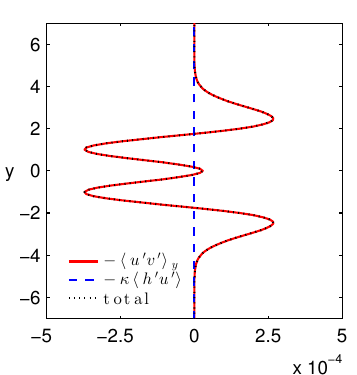} 
\end{center}
\caption{The effective zonal forcing terms $- \langle u' v' \rangle_y$ and $- \kappa \langle h' u' \rangle$ on the right hand side of (\ref{Umean}), and their sum, for (a) dissipation solely by radiative relaxation ($\kappa = 0.001, \gamma = 0$) and (b) dissipation solely by friction ($\kappa = 0, \gamma =0.001$). All terms are calculated for $n=1$, $k=1$, and $a=1$ for scaling. Panel (b) has the sum equal to the first term as $\kappa=0$.}
\label{fig:ReynoldsUone}


\begin{center}
(a) \includegraphics{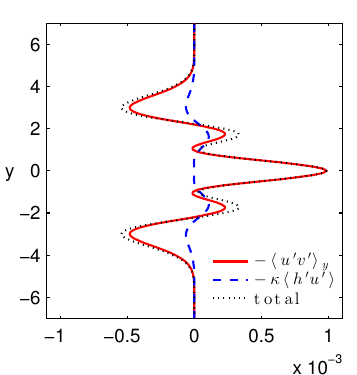} 
(b) \includegraphics{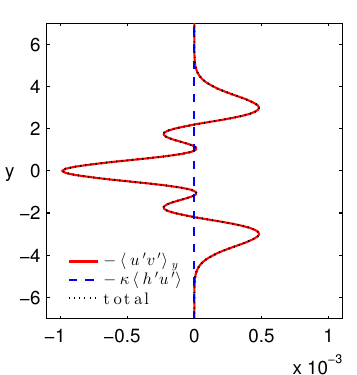} 
\end{center}
\caption{The effective zonal forcing terms $- \langle u' v' \rangle_y$ and $- \kappa \langle h' u' \rangle$ on the right hand side of \Ex{Umean}, and their sum, for (a) dissipation solely by radiative relaxation ($\kappa = 0.001, \gamma = 0$) and (b) dissipation solely by friction ($\kappa = 0, \gamma =0.001$). All terms are calculated for $n=2$, $k=1$, and $a=1$ for scaling. Panel (b) has the sum equal to the first term as $\kappa=0$.}
\label{fig:ReynoldsUtwo}

\begin{center}
(a) \includegraphics{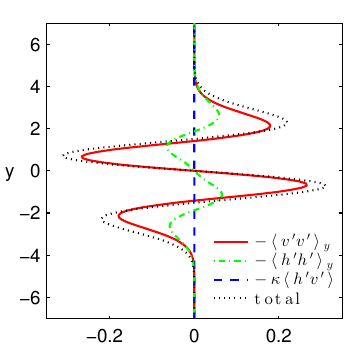} 
(b) \includegraphics{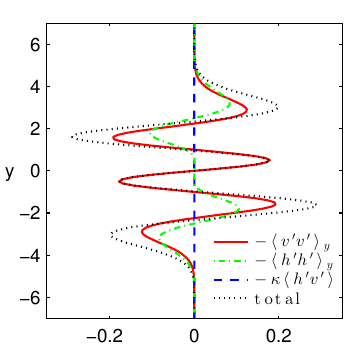} 
\end{center}
\caption{The effective meridional forcing terms $- \tfrac{1}{2} \langle h'^2 \rangle_y$, $- \langle v' v' \rangle_y$ and $- \kappa \langle h' u' \rangle$ on the right hand side of \Ex{Vmean}, and their sum, for dissipation solely by radiative relaxation ($\kappa = 0.001, \gamma = 0$) and $a=1$. All terms are calculated for $k=1$ and either (a) $n=1$ or (b) $n=2$. The bolus velocity term $- \kappa \langle h' u' \rangle$ is much smaller than the other two terms, which are not proportional to $\kappa$.}
\label{fig:ReynoldsVonetwo}

\end{figure}

Figures~\ref{fig:ReynoldsUone} and \ref{fig:ReynoldsUtwo} show the two
quantities $- \langle u' v' \rangle_y$ and $- \kappa \langle h' u'
\rangle$ on the right hand side of (\ref{Umean}), and their sum, for
waves with $k=1$ and $n \in \{1,2 \}$ for dissipation by either
Rayleigh friction ($\gamma > 0$, $\kappa=0$) or radiative relaxation
($\gamma = 0$, $\kappa>0$). The plots show the quantities
divided by $a^2$, or equivalently correspond to setting $a=1$.
The Reynolds stress $\langle u' v'
\rangle$ vanishes for undamped waves, since $\freq$, $A$, and
$\hat{v}$ are then all real, while $\hat{u}$ and $\hat{h}$ are purely
imaginary. The bolus velocity term $- \kappa \langle h' u' \rangle$ is
explicitly proportional to $\kappa$, and so vanishes for $\kappa =
0$. 

A perturbative analysis similar to that in \S\ref{subsec:Rossbywaves_h} shows that the Reynolds stress
$\langle u' v' \rangle$ is proportional to $\kappa - \gamma$ for small $\kappa$ and $\gamma$. This explains
the antisymmetry visible when comparing panel (a) with $\kappa > 0$, $\gamma=0$ panel (b) with $\kappa = 0$, $\gamma > 0$
in each of figures \ref{fig:ReynoldsUone} and \ref{fig:ReynoldsUtwo}.
Increasing $k$ while holding $n$, $\kappa$, $\gamma$ fixed
increases the magnitude of $\langle u'v' \rangle$ without changing its
general shape, as does increasing $\kappa$ or $\gamma$ while holding
$k$ and $n$ fixed. 

The Reynolds stress convergence is close to zero at the equator when $n$ is odd, leaving just the contribution from the bolus
velocity visible in figure~\ref{fig:ReynoldsUone}(a). However, the Reynolds stress convergence has a local extremum on the equator
when $n$ is even. Its sign for $n=2$ is such as to produce eastward acceleration at the equator under radiative relaxation (figure \ref{fig:ReynoldsUtwo}(a)), and an westward acceleration at the equator under Rayleigh friction (figure \ref{fig:ReynoldsUtwo}(b)). We can infer the sign of the acceleration
directly from the right hand side of \Ex{Umean} on the equator. Since $y=0$, the Coriolis contribution from the mean meridional velocity
$\langle v \rangle^*$ does not contribute to the left hand side of \Ex{Umean}. This leaves the ordinary differential equation
\begin{equation}
\dd{}{t} \langle u \rangle^{*} \big|_{y=0} + \gamma  \langle u \rangle^{*} \big|_{y=0} = -  \mathrm{e}^{- \sigma t} \left( \langle u' v' \rangle_y + \kappa \langle h' u' \rangle \right) \big|_{y=0,t=0}, \label{UdecayODE}
\end{equation}
where $\sigma = - 2 \Imag \, \omega$ is twice the decay rate of the Rossby wave calculated in \S\ref{subsec:Rossbywaves_h}. The quadratic
products on the right hand side of \Ex{Umean} thus decay in proportion to $\exp(-\sigma t)$. The solution of \Ex{UdecayODE} for $\gamma \ne \sigma$ is
\begin{equation}
 \langle u \rangle^{*} \big|_{y=0} = \left[\frac{\mathrm{e}^{-\gamma t} - \mathrm{e}^{-\sigma t}}{\sigma-\gamma} \right]
\left( \langle u' v' \rangle_y + \kappa \langle h' u' \rangle \right) \big|_{y=0,t=0}, \label{UdecayODEsoln}
\end{equation}
The time-dependent coefficient in square brackets $[\cdots]$ is always positive for $t>0$, so the direction of the zonal mean
flow $\langle u \rangle^{*}$ at the equator is set by the sign of the combined Reynolds stress and bolus velocity term
evaluated at $y=0$ and $t=0$. This is consistent with the directions of the equatorial
jets found in numerical simulations for different rates of Rayleigh friction and radiative relaxation \citep{ScottPolvani07, ScottPolvani08, WarnefordDellar14tsw}.

However, away from the equator one would need to compute the solution
to the full response problem coupling $\langle h \rangle$, $\langle u
\rangle^*$, $\langle v \rangle^*$ as three functions of $y$ and $t$ to find the mean flow produced by
the wave sources on the right hand sides of \Ex{Umean} and \Ex{Vmean}.
Figure~\ref{fig:ReynoldsVonetwo} shows the three terms $- \tfrac{1}{2} \langle h'^2 \rangle_y$, $- \langle v' v' \rangle_y$ and $- \kappa \langle h' u' \rangle$ on the right hand side of \Ex{Vmean}, and their sum, for $k=1$, $n \in \{1,2\}$ and dissipation solely by radiative relaxation ($\kappa = 0.001, \gamma = 0$). The first two terms are non-zero even with no dissipation, and typically much larger in magntitude than either the bolus terms or the Reynolds stress convergence in the zonal equation.

Omitting the $\kappa$ terms from the right hand sides of \Ex{eq:uflux} and \Ex{eq:vflux}  leads to a shallow water model that relaxes the thickeness field but conserves momentum. This model is no less justified than the model that supposes that $\kappa$ should not appear in the equation \Ex{eq:sp2} for evolving the velocity $\uv$. The contributions $\langle h' u' \rangle$ and $\langle h' v' \rangle$ from the bolus velocity would then be absent in the right hand sides of the analogues of \Ex{Umean} and \Ex{Vmean} for this model.
Figures \ref{fig:ReynoldsUone}(a) and \ref{fig:ReynoldsUtwo}(a) show that these contributions are in any case small compared with the contribution from the Reynolds stress convergence. This is consistent with the directions of the equatorial jets in the two models, with and without momentum conservation, showing the same behaviour in simulations \citep{WarnefordDellar14tsw}.

\subsection{Thermal shallow water equations} \label{subsec:mom_t}

We now apply the same approach to the dimensionless form of the unforced thermal shallow water equations (\ref{eq:tsw}) on an equatorial beta plane.
The mass and zonal momentum equations are 
\begin{equation} \label{HmeanT}
\langle h \rangle_{t} + \langle v \rangle^{*}_{y} = 0,
\end{equation}
and 
\begin{equation}
\langle u \rangle^{*}_t - \tfrac{1}{2} y \langle v \rangle^{*} + \gamma \langle u \rangle^{*}
= - \langle u' v' \rangle_y. \label{UmeanT} 
\end{equation}
There are no $\kappa$ terms in these equations, as radiative relaxation conserves mass and momentum in our thermal shallow
water model. The temperature $\Theta$ also does not appear in these equations, because the zonal pressure gradient disappears
under zonal averaging.

To calculate the thermal contributions to the other equations, it is convenient to decompose the temperature as
\begin{equation}
\Theta = 1 + \thetat = 1 + \underbrace{\thetat'}_{\Ord(a)} + \underbrace{\langle \thetat \rangle}_{\Ord(a^2)}.
\end{equation}
This decomposition eliminates $\Ord(1)$ terms proportional to the reference temperature $\Theta_0$, which is scaled to unity by the
nondimensionalisation in \S\ref{subsec:Rossbywaves_t}. Although $\thetat' = \Theta'$, we use $\thetat'$ below so that all equations
are written using $\thetat$.

The meridional momentum equation is
\begin{equation}
\left( h v \right)_{t} + \left( h u v \right)_{x} + \left(h v^2 + \tfrac{1}{2} h^{2} (1 + \thetat) \right)_{y} + \tfrac{1}{2} y h u = -\gamma h v, 
\label{VeqnT}
\end{equation}
and the conservation form of the  temperature equation is
\begin{equation}
(h \thetat)_t + (h u \thetat)_x +  (h v \thetat)_y = - \kappa \left( h^2(1+\thetat) - h \right). \label{TeqnT}
\end{equation}
The unit term from the decomposition $\Theta = 1 + \thetat$ disappears
from the left hand side of \Ex{TeqnT} because $h$ satisfies the mass conservation equation
with no source terms. However, the combination $h^2 (1+\thetat)$ appears
in both the pressure gradient in \Ex{VeqnT} and the radiative
relaxation term in \Ex{TeqnT}. We may reduce this expression to a quadratic product by writing
\begin{equation}
h^2 \thetat = h \thetat + h'\thetat' + \Ord(a^3).
\end{equation}

The zonal mean of the temperature equation may thus be written as
\begin{equation}
\langle \thetat \rangle^*_t + \kappa \left( \langle \thetat \rangle^* + \langle h \rangle - 1 \right) = - \langle v' \thetat' \rangle_y - \kappa \left( \langle h' \thetat \rangle + \langle h'^2 \rangle\right). \label{TmeanT}
\end{equation}
As before for the velocity components, we use $\langle \thetat \rangle^*$ instead of  $\langle \thetat \rangle$ to absorb a contribution from $\langle \thetat' h'\rangle$ in the time derivative.
As $\thetat = \Ord(a)$ under our decomposition, we may neglect factors of $\langle h \rangle = 1 + \Ord(a^2)$ that we could
not neglect for $\langle \Theta \rangle^*$. Similarly, the zonal mean of the meridional momentum equation is
\begin{equation}
\langle v \rangle^{*}_t + \tfrac{1}{2} y \langle u \rangle^{*} + \gamma \langle v \rangle^{*} 
+ \langle h \rangle_y + \tfrac{1}{2}  \langle \thetat \rangle^{*}_y
=  - \langle v' v' \rangle_y - \tfrac{1}{2} \langle h'^2 \rangle_y - \tfrac{1}{2} \langle h' \thetat' \rangle_y,  \label{VmeanT}
\end{equation}
so we again have four equations for evolving $\langle h
\rangle$, $\langle u \rangle^{*}$, $\langle u \rangle^{*}$, $\langle
\thetat \rangle^*$ with linear left hand sides. The right hand sides are again
zonal means of quadratic products of the fluctuating quantities $h'$, $u'$, $v'$, $\thetat'$.

The meridional velocity perturbation for an equatorial Rossby wave is again $v' = \alpha \, \Real \{ \hat{v}(y) \exp(\mathrm{i} (k x-\freq t)) \}$, with $\hat{v}(y)$ defined by \Ex{TrapHermite} and (\ref{eq:AB_thermal}). The 
corresponding perturbations to the thickness, temperature, and zonal velocity take the same functional forms, with
\begin{align} \label{eq:uhat_t}
\hat{h}(y) &= - \mathrm{i} \left( \frac{\mathrm{i} k  \hat{u}  +  \mbox{d}\hat{v} / \mbox{d} y}{\freq} \right), \quad  \quad \hat{\Theta}(y) = - \mathrm{i} \left( \frac{\kappa \hat{h} }{\freq + \mathrm{i} \kappa} \right), \nonumber \\
\hat{u}(y) &= \mathrm{i} \left( \frac{ k \left( 1 + \mathrm{i} \kappa / \left(2 \freq \right) \right) \mbox{d}\hat{v} / \mbox{d} y - \tfrac{1}{2} y \left( \freq + \mathrm{i} \kappa \right) \hat{v}}{k^2 \left( 1 + \mathrm{i} \kappa / \left( 2 \freq \right) \right) - \left( \freq + \mathrm{i} \gamma \right) \left( \freq + \mathrm{i} \kappa \right)} \right),
\end{align}
for $\freq \neq 0$ and $\freq \ne - \mathrm{i} \kappa$. The vanishing denominator in the expression for $\hat{u}$
when $k^2 = \left( \freq + \mathrm{i} \gamma \right) \left( \freq + \mathrm{i} \kappa \right) / \left( 1 + \mathrm{i} \kappa / \left( 2 \freq \right) \right) $ again gives the dispersion relation for the equatorially trapped Kelvin waves.  We choose the normalisation constant $\alpha$ so that the disturbances $u'$, $v'$, $h'$ and $\thetat'$ satisfy
\begin{equation}
\frac{1}{2} \int \langle u'^2 \rangle + \langle v'^2 \rangle + \langle (h'+\tfrac{1}{2} \thetat')^2 \rangle \, \mbox{d} y = a^2. \label{eq:PseudoTSW}
\end{equation}
The left hand side is the dimensionless quadratic approximation for small amplitude disturbances to the pseudoenergy
\begin{equation}
\frac{1}{2} \int h \left( u^2 + v^2 \right) + \left( h \sqrt{\Theta} - h_0 \sqrt{\Theta_0} \right)^{2} \, \mbox{d} x \, \mbox{d} y \label{eq:PseudoTSWfull}
\end{equation}
for finite amplitude disturbances from a rest state of constant depth $h_0$ and temperature $\Theta_0$ in
the thermal shallow water equations \citep{Ripa95,Roed97,WarnefordDellarQGTSW}


\begin{figure}
\begin{center}
(a) \includegraphics{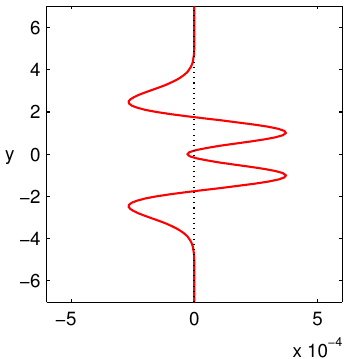} 
(b) \includegraphics{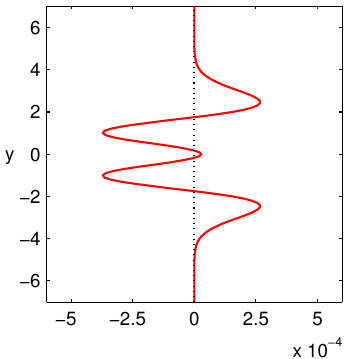} 
\end{center}
\caption{The effective zonal force $- \langle u' v' \rangle_y$ on the right hand side of (\ref{UmeanT}) for $k=1$, $n=1$, and $a=1$ for scaling. Panel
(a) shows dissipation solely by radiative relaxation ($\kappa = 0.002, \gamma = 0$) and panel (b) shows dissipation solely by friction ($\kappa = 0, \gamma =0.001$). The zero line is shown dotted. The ratio of $\kappa$ to $\gamma$ reflects the symmetry of the dispersion relation in \S\ref{subsec:Rossbywaves_t}.}
\label{fig:TReynoldsUone}


\begin{center}
(a) \includegraphics{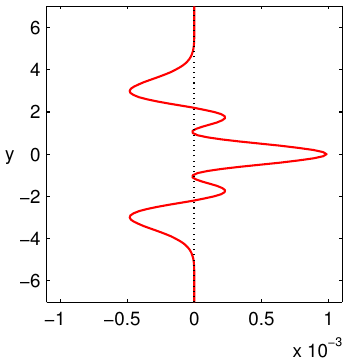} 
(b) \includegraphics{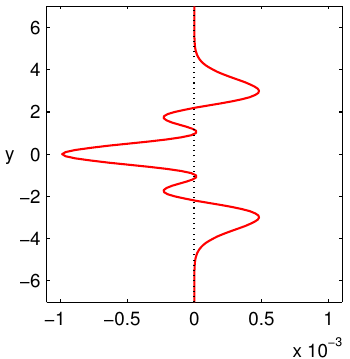} 
\end{center}
\caption{The effective zonal force $- \langle u' v' \rangle_y$ on the right hand side of (\ref{UmeanT}) for $k=1$, $n=2$, and $a=1$ for scaling. Panel
(a) shows dissipation solely by radiative relaxation ($\kappa = 0.002, \gamma = 0$) and panel (b) shows dissipation solely by friction ($\kappa = 0, \gamma =0.001$). The zero line is shown dotted. The ratio of $\kappa$ to $\gamma$ reflects the symmetry of the dispersion relation in \S\ref{subsec:Rossbywaves_t}. }
\label{fig:TReynoldsUtwo}

\end{figure}

\begin{figure}
\begin{center}
(a) \includegraphics{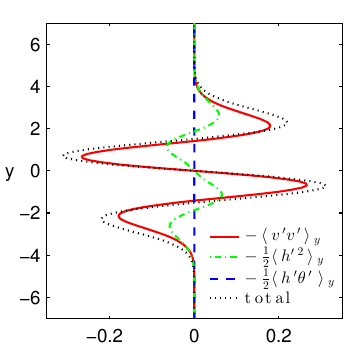} 
(b) \includegraphics{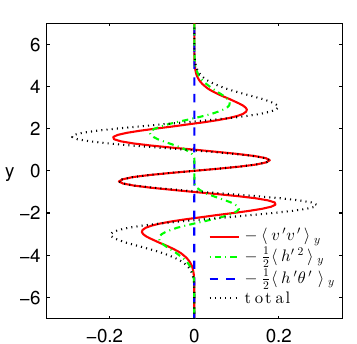} 
\end{center}
\caption{The effective meridional forcing terms $- \langle v' v'' \rangle_y$, $- \tfrac{1}{2} \langle h'^2 \rangle_y$, $- \tfrac{1}{2} \langle h'\thetat' \rangle_y$ on the right hand side of \Ex{VmeanT}, and their sum, for dissipation solely by radiative relaxation ($\kappa = 0.002, \gamma = 0$). All terms are calculated for $k=1$, $a=1$ for scaling, and either (a) $n=1$ or (b) $n=2$.}
\label{fig:TReynoldsVonetwo}
\end{figure}

\begin{figure}
\begin{center}
(a) \includegraphics{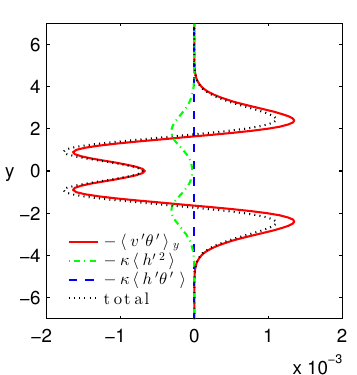} 
(b) \includegraphics{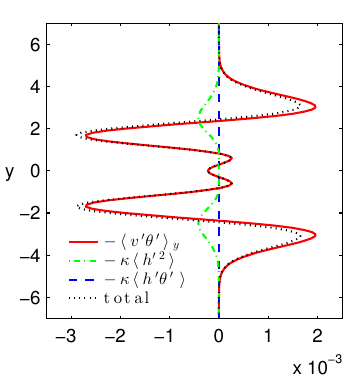} 
\end{center}
\caption{The effective thermal terms $- \langle v' \thetat' \rangle_y$, $- \kappa \langle h' \thetat' \rangle$, $- \kappa\langle h'^2\rangle$ on the right hand side of \Ex{TmeanT}, and their sum, for dissipation solely by radiative relaxation ($\kappa = 0.002, \gamma = 0$). All terms are calculated for $k=1$, $a=1$ for scaling,  and either (a) $n=1$ or (b) $n=2$.}
\label{fig:TReynoldsTonetwo}
\end{figure}

Figures \ref{fig:TReynoldsUone} and \ref{fig:TReynoldsUtwo} show the Reynolds stress convergences on the right hand side of \Ex{UmeanT}. These all closely resemble those for the
previous shallow water equations with radiative relaxation of the thickness. Thus the mean flow at the equator ($y=0$) is almost identical to that for the previous model, provided $\kappa$ is rescaled by a factor of $2$. A perturbative analysis similar to that in \S\ref{subsec:Rossbywaves_h} shows that the Reynolds stress $\langle u' v' \rangle$ for this model is proportional to $\kappa/2 - \gamma$ for small $\kappa$ and $\gamma$.

Figure \ref{fig:TReynoldsVonetwo} shows the forcing terms in the right hand side of the mean meridional momentum equation
\Ex{VmeanT}. The dominant terms, $-\langle v'v' \rangle_y$ and $-\tfrac{1}{2} \langle h'^2 \rangle_y$ are
very similar to those in figure \ref{fig:ReynoldsVonetwo} for the shallow water model with relaxation of the thickness. The  thermal term $-\tfrac{1}{2} \langle h' \thetat' \rangle_y$ is much smaller.
Finally, figure \ref{fig:TReynoldsTonetwo} shows the forcing terms in the right hand side of the mean temperature equation \Ex{TmeanT}.
These all vanish for dissipation solely by friction ($\kappa = 0$), since the temperature perturbation $\hat \theta$ given by \Ex{eq:uhat_t} then vanishes, so they are only shown for $\kappa > 0$ and $\gamma=0$. As for the meridional velocity, the contribution from $\langle h' \thetat' \rangle$ is much smaller than the other contributions.

\subsection{Discussion}


Our simulation parameters imply a dimensionless radiative relaxation rate $\kappa = 7.88 \times 10^{-3}$ roughly 23 times larger than our dimensionless frictional rate $\gamma = 3.42 \times 10^{-4}$. In \S \ref{subsec:mom_h} we established that the Reynolds stress $\langle u'v' \rangle$ in the mean zonal momentum equation is proportional to $\kappa - \gamma$ (for thickness relaxation) or $\kappa/2-\gamma$ (for temperature relaxation) for small relaxation and friction rates. For $\kappa \gg \gamma$, our theory thus suggests that $\langle u \rangle^* > 0$ on the equator, consistent with the super-rotating equatorial jet in simulations. Conversely, for the purely frictional case ($\kappa=0$) our theory gives $\langle u \rangle^* < 0$ on the equator, again consistent with the sub-rotating equatorial jet in simulations. These conclusions hold for both shallow water models, with either the thickness or the temperature being relaxed. The Reynolds stresses in the two models are very similar for small relaxation and friction rates, provided one rescales the radiative relaxation rate $\kappa$ in the thermal model by a factor of 2, as in figures \ref{fig:TReynoldsUone}(a) and \ref{fig:TReynoldsUtwo}(a).


We have performed a similar analysis for decaying equatorially trapped
Kelvin, Yanai, and inertia-gravity waves in our two shallow water
models with radiative relaxation of the thickness or temperature. The
effective zonal forces due to these waves are at least an order of
magnitude smaller than those due to decaying Rossby waves of the same
pseudoenergy. Thus, as expected from our global QG simulations, these
other types of wave make no significant contribution to setting the
direction of the equatorial jet.

Our thermal shallow water model conserves both mass and momentum
in the absence of friction ($\gamma = 0$). This offers a simple demonstration
that there is no inconsistency between our calculations of positive (eastward) zonal accelerations at the
equator due to radiative dissipation of Rossby waves and the arguments of \S\ref{sect:trans} that Rossby waves should
carry westward pseudomomentum. The evolution equation for $u$ may be rewritten
in conservation form as
\begin{equation}
\partial_t \big( h (u-y^2/4) \big) + \partial_x \big( (hu (u-y^2/4) + \Theta h^2/2 \big) + \partial_y \big( hv (u-y^2/4) \big) = 0. \label{UconsY}
\end{equation}
The combination $u - y^2/4$ corresponds to the zonal velocity seen in an inertial frame, such that
$\pd_x v - \pd_y (u - y^2/4) = y/2 + \pd_x v - \pd_y u$ is the absolute vorticity \citep{Ripa93}. The combination $h(u - y^2/4$), sometimes called absolute zonal momentum, 
is the beta-plane analogue of azimuthal angular momentum in spherical geometry.

 The zonal mean of \Ex{UconsY} is
\begin{equation}
\partial_t \big\langle h (u-y^2/4) \big\rangle  + \partial_y \big\langle hv (u-y^2/4) \big\rangle = 0,
\end{equation}
which becomes
\begin{equation}
\partial_t \left( \langle u \rangle^* - \tfrac{1}{4} y^2 \langle h \rangle \right)
+ \partial_y \left( \langle u' v' \rangle  -  \tfrac{1}{4} y^2 \langle h v \rangle  \right) = 0,
\label{ubarcons}
\end{equation}
correct to $\Ord(a^2)$. Integrating \eqref{ubarcons} over $y$ implies that the
global absolute zonal momentum 
\begin{equation}
\mathcal{M} = \int_{-\infty}^\infty \mbox{d} y \, \langle u \rangle^* -  \tfrac{1}{4} y^2 (\langle h \rangle -1) \label{Mdef}
\end{equation}
is constant in time. We have added the time-independent term $y^2/4$ to the integrand defining $\mathcal{M}$ so that this integral converges when $\langle h \rangle \to 1$ as $y \to \pm \infty$.
Evaluating $\mathcal{M}$ at $t=0$ for the equatorially trapped Rossby wave solution from \S \ref{subsec:Rossbywaves_t} for which $\langle h \rangle = 1$ and $\langle u \rangle = 0$ initially, we obtain
\begin{equation} \label{eq:M_t_0}
\mathcal{M} \bigg|_{t=0} = \int_{-\infty}^{\infty} \mbox{d} y \, \tfrac{1}{2} \Real \{ \hat{h}(y) \hat{u}(y)^{*} \}  < 0,
\end{equation}
where $\hat{h}(y)$ and $\hat{u}(y)$ are defined in (\ref{eq:uhat_t}), and $\hat{v}(y)$ is defined by \Ex{TrapHermite} and (\ref{eq:AB_thermal}).
This negative sign agrees with the sign of the total zonal momentum due to a Rossby wave, as described in \S\ref{sect:trans}.

The mass-weighted mean zonal velocity $\langle u \rangle^*$ generated by the decaying
Rossby wave may be positive (eastward) at the equator, but the meridional integral of $\langle u \rangle^* - \tfrac{1}{4} y^2 (\langle h \rangle -1)$ that defines $\mathcal{M}$
is still negative (westward). This is all that is required to
reconcile conservation of zonal angular momentum with the requirement
that Rossby waves should carry westward zonal pseudomomentum, as
established concretely by \eqref{eq:M_t_0}.

\section{Conclusions}

Numerical simulations of shallow water equations on a sphere with
isotropic random forcing reliably produce a mix of coherent
vortices and alternating zonal jets. These simulations typically
include Rayleigh friction to absorb the gradual inverse cascade of
energy past the Rhines scale, since the zonal jets would otherwise
break up into domain-sized coherent vortices after very long
times. Such simulations invariably produce sub-rotating equatorial
jets when run with Jovian parameters
\citep{VasavadaShowman05}. However, simulations that model radiative
effects by relaxing the thickness field produce a super-rotating
equatorial jet \citep{ScottPolvani07, ScottPolvani08}, as do
simulations of the thermal shallow water equations with a
radiative relaxation term in the temperature equation \citep[see \S
\ref{sect.num} and][]{WarnefordDellar14tsw}.

We have provided a possible explanation for the different directions
of the equatorial jets in different shallow water models with
dissipation by Rayleigh friction or Newtonian radiative relaxation. We
formulated evolution equations for the zonal mean thickness $\langle h
\rangle$, and the thickness-weighted zonal mean quantities $\langle u
\rangle^*$, $\langle v \rangle^*$, $\langle \theta \rangle^*$.  These
equations contain effective forces due to zonal means of products of
fluctuations, such as the Reynolds stress convergence $\langle u'v'
\rangle_y$ in the evolution equation for $\langle v \rangle^*$. We
evaluated these effective forces for an equatorially trapped Rossby
waves superimposed on a background rest state on an equatorial beta
plane. Dissipation changes the usual phase relations between the
perturbations $u'$, $v'$, $h'$, and permits a nonzero Reynolds stress
$\langle u'v' \rangle$ that transports zonal momentum in the
meridional direction. This Reynolds stress is proportional to $\kappa
- \gamma$ for the shallow water model with relaxation of the
thickness, and to $\kappa/2 - \gamma$ for the thermal shallow water
model with relaxation of the temperature. The Reynolds stress divergence
$ \langle u'v' \rangle_y$ is the sole forcing term in the momentum-conserving
shallow water models, and the larger forcing term in the \cite{ScottPolvani07, ScottPolvani08}
model with radiative relaxation of both thickness and momentum. In all cases, radiative
relaxation produces a Reynolds stress with the opposite sign to that due to Rayleigh friction.
This suffices to determine that the zonal mean flow $\langle u \rangle^*$ on the equator created
by the decaying Rossby wave with $n=2$ is eastward (super-rotating) for dissipation by radiative relaxation,
and westward (sub-rotating) for dissipation by Rayleigh friction.

Our work relies upon several major simplifying assumptions. We
considered trapped equatorial Rossby waves on a background rest state,
which allowed us to use the known analytical expressions for these
waves in terms of Hermite polynomials. A more complete theory would
calculate the Rossby wave spectrum supported by a zonally symmetric
background flow. Secondly, we only calculated the effective forces in the evolution
equations for the thickness-weighted zonal mean quantities $\langle u \rangle^*$,
 $\langle v \rangle^*$, $\langle \tilde \theta \rangle^*$. This is only sufficient
to determine the evolution of $\langle u \rangle^*$ at the equator. Elsewhere, one
would need to solve the system of three or four coupled equations to determine the
Coriolis contribution to $\langle u \rangle^*$ from $\langle v \rangle^*$. However, it
is worth noting that the latitude-dependence of the Reynolds stress divergence $\langle u'v' \rangle_y$ for the $n=1$
equatorially trapped Rossby wave, with two local maxima around $y = \pm 1$
in figures \ref{fig:ReynoldsUone}(a) and  \ref{fig:TReynoldsUone}(a), 
is suggestive of the
presence of two ``horns'', peaks of maximal eastward velocity, at latitudes
around $\pm 7 ^\circ$ near the edges of the broad Jovian equatorial jet, and a slightly smaller
velocity precisely on the equator (see figure \ref{fig:ubardata}).

Finally, we assumed that the waves are
excited by sources outside the equatorial region, and neglected the
effect of this excitation on the zonal mean flow. Some support for
this second assumption is offered by simulations of the quasilinear version of
the \cite{ScottPolvani07,ScottPolvani08} model in spherical geometry
by \cite{SaitoIshioka14} which show that the zonal mean zonal
acceleration due to the ensemble of Rossby waves excited by random
forcing of the wavenumbers in the annulus described in
\S\ref{sect.num}, and decaying by either radiative relaxation or friction, produce
an equatorial jet in the direction predicted by our arguments, and
with an amplitude consistent at early times with their fully nonlinear
simulations. They also gave an elegant argument to determine the
direction of the zonal acceleration from the tilting of the constant phase
lines in the separable Rossby wave solutions, as previously demonstrated by \cite{AndrewsMcIntyre76b} for stratified
fluids, and later adapted to shallow water models by \cite{Mofjeld81} and \cite{YamagataPhilander85}.
Although our work
falls short of a full wave-mean flow interaction theory, we believe it offers at least a step towards a theoretical
explanation of the origins of equatorial super- or sub-rotation
in shallow water  simulations of Jovian atmospheres.

\begin{acknowledgements}

  We thank Michael McIntyre for very detailed and stimulating referee
  reports that greatly improved our manuscript, and Rick Salmon and Ted Johnson for useful conversations.
  This work was supported by the Engineering and Physical Sciences
  Research Council through a Doctoral Training Grant award to
  E.S.W. and an Advanced Research Fellowship to P.J.D. [grant number
  EP/E054625/1].  The computations employed the University of Oxford's
  Advanced Research Computing facilities \citep{ARC}, and the Emerald
  GPU-accelerated High Performance Computer made available by the
  Science and Engineering South Consortium in partnership with the
  Science and Technology Facilities Council's Rutherford--Appleton
  Laboratory.

\end{acknowledgements}


\begin{thebibliography}{75}
\expandafter\ifx\csname natexlab\endcsname\relax\def\natexlab#1{#1}\fi

\bibitem[Andrews \& McIntyre(1976{\natexlab{{\em a\/}}})]{AndrewsMcIntyre76a}
{\sc Andrews, D. \& McIntyre, M.} 1976{\natexlab{{\em a\/}}} {Planetary waves
  in horizontal and vertical shear: The generalized Eliassen--Palm relation and
  the mean zonal acceleration}. {\em J. Atmos. Sci.\/} {\bf 33}, 2031--2048.

\bibitem[Andrews \& McIntyre(1976{\natexlab{{\em b\/}}})]{AndrewsMcIntyre76b}
{\sc Andrews, D. \& McIntyre, M.} 1976{\natexlab{{\em b\/}}} Planetary waves in
  horizontal and vertical shear: Asymptotic theory for equatorial waves in weak
  shear. {\em J. Atmos. Sci.\/} {\bf 33}, 2049--2053.

\bibitem[Andrews \& McIntyre(1978)]{AndrewsMcIntyre78c}
{\sc Andrews, D.~G. \& McIntyre, M.~E.} 1978 {Generalized Eliassen--Palm and
  Charney--Drazin theorems for waves on axisymmetric mean flows in compressible
  atmospheres}. {\em J. Atmos. Sci.\/} {\bf 35}, 175--185.

\bibitem[Beebe(1994)]{Beebe1994}
{\sc Beebe, R.} 1994 Characteristic zonal winds and long-lived vortices in the
  atmospheres of the outer planets. {\em Chaos\/} {\bf 4}, 113--122.

\bibitem[B\"uhler(2000)]{Buhler00}
{\sc B\"uhler, O.} 2000 On the vorticity transport due to dissipating or
  breaking waves in shallow-water flow. {\em J. Fluid Mech.\/} {\bf 407},
  235--263.

\bibitem[B\"uhler(2014)]{Buhler14book}
{\sc B\"uhler, O.} 2014 {\em {Waves and Mean Flows}\/}, 2nd edn. Cambridge:
  Cambridge University Press.

\bibitem[Charney(1949)]{Charney49}
{\sc Charney, J.~G.} 1949 On a physical basis for numerical prediction of
  large-scale motions in the atmosphere. {\em J. Atmos. Sci.\/} {\bf 6},
  372--385.

\bibitem[Cho \& Polvani(1996{\natexlab{{\em a\/}}})]{ChoPolvani96a}
{\sc Cho, J. Y.-K. \& Polvani, L.~M.} 1996{\natexlab{{\em a\/}}} The emergence
  of jets and vortices in freely evolving, shallow-water turbulence on a
  sphere. {\em Phys. Fluids\/} {\bf 8}, 1531--1552.

\bibitem[Cho \& Polvani(1996{\natexlab{{\em b\/}}})]{ChoPolvani96b}
{\sc Cho, J. Y.-K. \& Polvani, L.~M.} 1996{\natexlab{{\em b\/}}} The
  morphogenesis of bands and zonal winds in the atmospheres on the giant outer
  planets. {\em Science\/} {\bf 273}, 335--337.

\bibitem[Daley(1983)]{Daley83}
{\sc Daley, R.} 1983 Linear non-divergent mass-wind laws on the sphere. {\em
  Tellus A\/} {\bf 35A}, 17--27.

\bibitem[Dickinson(1969)]{Dickinson69a}
{\sc Dickinson, R.~E.} 1969 Theory of planetary wave-zonal flow interaction.
  {\em J. Atmos. Sci.\/} {\bf 26}, 73--81.

\bibitem[Dowling(1995)]{dowling1995estimate}
{\sc Dowling, T.~E.} 1995 Estimate of {Jupiter's} deep zonal-wind profile from
  {Shoemaker--Levy} 9 data and {Arnol'd's} second stability criterion. {\em
  Icarus\/} {\bf 117}, 439--442.

\bibitem[Dowling \& Ingersoll(1989)]{DowlingIngersoll89}
{\sc Dowling, T.~E. \& Ingersoll, A.~P.} 1989 {Jupiter's Great Red Spot as a
  shallow water system}. {\em J. Atmos. Sci.\/} {\bf 46}, 3256--3278.

\bibitem[Gill(1982)]{gill82}
{\sc Gill, A.~E.} 1982 {\em Atmosphere Ocean Dynamics\/}. New York: Academic
  Press.

\bibitem[Green(1970)]{Green70}
{\sc Green, J. S.~A.} 1970 Transfer properties of the large-scale eddies and
  the general circulation of the atmosphere. {\em Q. J. R. Meteorol. Soc.\/}
  {\bf 96}, 157--185.

\bibitem[Haynes \& McIntyre(1990)]{HaynesMcIntyre90}
{\sc Haynes, P.~H. \& McIntyre, M.~E.} 1990 On the conservation and
  impermeability theorems for potential vorticity. {\em J. Atmos. Sci.\/} {\bf
  47}, 2021--2031.

\bibitem[Held(2000)]{Held00whoi}
{\sc Held, I.~M.} 2000 {The General Circulation of the Atmosphere:
  Superrotation}. Lectures presented at the 2000 Geophysical Fluid Dynamics
  Summer Program, Woods Hole Oceanographic Institution, Woods Hole, MA,
  available from \url{http://www.whoi.edu/page.do?pid=13076}.

\bibitem[Hirst(1986)]{Hirst86}
{\sc Hirst, A.~C.} 1986 Unstable and damped equatorial modes in simple coupled
  ocean-atmosphere models. {\em J. Atmos. Sci.\/} {\bf 43}, 606--632.

\bibitem[Hou \& Li(2007)]{hou2007computing}
{\sc Hou, T.~Y. \& Li, R.} 2007 Computing nearly singular solutions using
  pseudo-spectral methods. {\em J. Comp. Phys.\/} {\bf 226}, 379--397.

\bibitem[Hupca {\em et~al.\/}(2012)Hupca, Falcou, Grigori \&
  Stompor]{HupcaFalcouGrigoriStompor12}
{\sc Hupca, I.~O., Falcou, J., Grigori, L. \& Stompor, R.} 2012 {Spherical
  harmonic transform with GPUs}. {\em Lecture Notes in Computer Science\/} {\bf
  7155}, 355--366.

\bibitem[Iacono {\em et~al.\/}(1999{\natexlab{{\em a\/}}})Iacono, Struglia \&
  Ronchi]{iacono1999spontaneous}
{\sc Iacono, R., Struglia, M.~V. \& Ronchi, C.} 1999{\natexlab{{\em a\/}}}
  Spontaneous formation of equatorial jets in freely decaying shallow water
  turbulence. {\em Phys. Fluids\/} {\bf 11}, 1272--1274.

\bibitem[Iacono {\em et~al.\/}(1999{\natexlab{{\em b\/}}})Iacono, Struglia,
  Ronchi \& Nicastro]{iacono1999high}
{\sc Iacono, R., Struglia, M.~V., Ronchi, C. \& Nicastro, S.}
  1999{\natexlab{{\em b\/}}} High-resolution simulations of freely decaying
  shallow-water turbulence on a rotating sphere. {\em Il Nuovo Cimento C\/}
  {\bf 22}, 813--821.

\bibitem[Ingersoll(1990)]{Ingersoll1990}
{\sc Ingersoll, A.~P.} 1990 Atmospheric dynamics of the outer planets. {\em
  Science\/} {\bf 248}, 308--315.

\bibitem[Ingersoll {\em et~al.\/}(2007)Ingersoll, Dowling, Gierasch, Orton,
  Read, S\'anchez-Lavega, Showman, Simon-Miller \&
  Vasavada]{IngersollDowlingGieraschEtAl04}
{\sc Ingersoll, A.~P., Dowling, T.~E., Gierasch, P.~J., Orton, G.~S., Read,
  P.~L., S\'anchez-Lavega, A., Showman, A.~P., Simon-Miller, A.~A. \& Vasavada,
  A.~R.} 2007 Dynamics of {Jupiter's} atmosphere. In {\em Jupiter: The Planet,
  Satellites and Magnetosphere\/} (ed. F.~Bagenal, T.~E. Dowling \& W.~B.
  McKinnon), pp. 105--128. Cambridge: Cambridge University Press.

\bibitem[Juckes(1989)]{Juckes89}
{\sc Juckes, M.} 1989 A shallow water model of the winter stratosphere. {\em J.
  Atmos. Sci.\/} {\bf 46}, 2934--2956.

\bibitem[Khouider {\em et~al.\/}(2013)Khouider, Majda \&
  Stechmann]{KhouiderMajdaStechmann13}
{\sc Khouider, B., Majda, A.~J. \& Stechmann, S.~N.} 2013 {Climate science in
  the tropics: waves, vortices and PDEs}. {\em Nonlinearity\/} {\bf 26},
  R1--R68.

\bibitem[Lavoie(1972)]{Lavoie72}
{\sc Lavoie, R.~L.} 1972 A mesoscale numerical model of lake-effect storms.
  {\em J. Atmos. Sci.\/} {\bf 29}, 1025--1040.

\bibitem[Limaye(1986)]{limaye1986jupiter}
{\sc Limaye, S.~S.} 1986 Jupiter: {New} estimates of the mean zonal flow at the
  cloud level. {\em Icarus\/} {\bf 65}, 335--352.

\bibitem[Liu \& Schneider(2010)]{LiuSchneider10}
{\sc Liu, J. \& Schneider, T.} 2010 Mechanisms of jet formation on the giant
  planets. {\em J. Atmos. Sci.\/} {\bf 67}, 3652--3672.

\bibitem[Majda \& Klein(2003)]{MajdaKlein03}
{\sc Majda, A.~J. \& Klein, R.} 2003 Systematic multiscale models for the
  tropics. {\em J. Atmos. Sci.\/} {\bf 60}, 393--408.

\bibitem[Marcus(1988)]{Marcus88}
{\sc Marcus, P.~S.} 1988 {Numerical simulation of Jupiter's Great Red Spot}.
  {\em Nature\/} {\bf 331}, 693--696.

\bibitem[Matsuno(1966)]{Matsuno66}
{\sc Matsuno, T.} 1966 Quasi-geostrophic motions in the equatorial area. {\em
  J. Meteorol. Soc. Japan\/} {\bf 44}, 25--43.

\bibitem[Matsuno(1970)]{Matsuno70}
{\sc Matsuno, T.} 1970 Vertical propagation of stationary planetary waves in
  the winter northern hemisphere. {\em J. Atmos. Sci.\/} {\bf 27}, 871--883.

\bibitem[Matsuno(1971)]{Matsuno71}
{\sc Matsuno, T.} 1971 A dynamical model of the stratospheric sudden warming.
  {\em J. Atmos. Sci.\/} {\bf 28}, 1479--1494.

\bibitem[McCreary(1985)]{McCreary85}
{\sc McCreary, J.~P.} 1985 Modeling equatorial ocean circulation. {\em Annu.
  Rev. Fluid Mech.\/} {\bf 17}, 359--409.

\bibitem[McCreary {\em et~al.\/}(1991)McCreary, Fukamachi \&
  Kundu]{mccreary1991numerical}
{\sc McCreary, J.~P., Fukamachi, Y. \& Kundu, P.~K.} 1991 A numerical
  investigation of jets and eddies near an eastern ocean boundary. {\em J.
  Geophys. Res.\/} {\bf 96}, 2515--2534.

\bibitem[McCreary \& Kundu(1988)]{mccreary1988numerical}
{\sc McCreary, J.~P. \& Kundu, P.~K.} 1988 A numerical investigation of the
  {Somali} {Current} during the {Southwest} {Monsoon}. {\em J. Marine Res.\/}
  {\bf 46}, 25--58.

\bibitem[McCreary \& Yu(1992)]{mccreary1992equatorial}
{\sc McCreary, J.~P. \& Yu, Z.} 1992 Equatorial dynamics in a $2
  \tfrac{1}{2}$-layer model. {\em Prog. Oceanogr.\/} {\bf 29}, 61--132.

\bibitem[McIntyre(1981)]{McIntyre81}
{\sc McIntyre, M.~E.} 1981 On the `wave momentum' myth. {\em J. Fluid Mech.\/}
  {\bf 106}, 331--347.

\bibitem[McIntyre \& Norton(1990)]{McIntyreNorton90}
{\sc McIntyre, M.~E. \& Norton, W.~A.} 1990 Dissipative wave-mean interactions
  and the transport of vorticity or potential vorticity. {\em J. Fluid Mech.\/}
  {\bf 212}, 403--435.

\bibitem[Mofjeld(1981)]{Mofjeld81}
{\sc Mofjeld, H.~O.} 1981 An analytic theory on how friction affects free
  internal waves in the equatorial waveguide. {\em J. Phys. Oceanogr.\/} {\bf
  11}, 1585--1590.

\bibitem[Obukhov(1949)]{Obukhov49}
{\sc Obukhov, A.~M.} 1949 On the problem of the geostrophic wind. {\em Izv.
  Akad. Nauk SSSR, Ser. Geograf. Geofiz.\/} {\bf 13}, 281--306.

\bibitem[Philander {\em et~al.\/}(1984)Philander, Yamagata \&
  Pacanowski]{PhilanderYamagataPacanowski84}
{\sc Philander, S. G.~H., Yamagata, T. \& Pacanowski, R.~C.} 1984 Unstable
  air-sea interactions in the tropics. {\em J. Atmos. Sci.\/} {\bf 41},
  604--613.

\bibitem[Polvani {\em et~al.\/}(1995)Polvani, Waugh \&
  Plumb]{PolvaniWaughPlumb95}
{\sc Polvani, L.~M., Waugh, D.~W. \& Plumb, R.~A.} 1995 On the subtropical edge
  of the stratospheric surf zone. {\em J. Atmos. Sci.\/} {\bf 52}, 1288--1309.

\bibitem[Porco {\em et~al.\/}(2003)Porco, West, McEwen, Del~Genio, Ingersoll,
  Thomas, Squyres, Dones, Murray, Johnson, Burns, Brahic, Neukum, Veverka,
  Barbara, Denk, Evans, Ferrier, Geissler, Helfenstein, Roatsch, Throop,
  Tiscareno \& Vasavada]{porco2003cassini}
{\sc Porco, C.~C., West, R.~A., McEwen, A., Del~Genio, A.~D., Ingersoll, A.~P.,
  Thomas, P., Squyres, S., Dones, L., Murray, C.~D., Johnson, T.~V., Burns,
  J.~A., Brahic, A., Neukum, G., Veverka, J., Barbara, J.~M., Denk, T., Evans,
  M., Ferrier, J.~J., Geissler, P., Helfenstein, P., Roatsch, T., Throop, H.,
  Tiscareno, M. \& Vasavada, A.~R.} 2003 Cassini imaging of {Jupiter's}
  atmosphere, satellites, and rings. {\em Science\/} {\bf 299}, 1541--1547.

\bibitem[{Rhines}(1975)]{Rhines75}
{\sc {Rhines}, P.~B.} 1975 {Waves and turbulence on a beta-plane}. {\em J.
  Fluid Mech.\/} {\bf 69}, 417--443.

\bibitem[Richards(2015)]{ARC}
{\sc Richards, A.} 2015 {University of Oxford Advanced Research Computing}.
  Zenodo document \url{doi:10.5281/zenodo.22558}.

\bibitem[Ripa(1993)]{Ripa93}
{\sc Ripa, P.} 1993 Conservation laws for primitive equations models with
  inhomogeneous layers. {\em Geophys. Astrophys. Fluid Dynam.\/} {\bf 70},
  85--111.

\bibitem[Ripa(1995)]{Ripa95}
{\sc Ripa, P.} 1995 On improving a one-layer ocean model with thermodynamics.
  {\em J. Fluid Mech.\/} {\bf 303}, 169--201.

\bibitem[Ripa(1996{\natexlab{{\em a\/}}})]{Ripa96b}
{\sc Ripa, P.} 1996{\natexlab{{\em a\/}}} Low frequency approximation of a
  vertically averaged ocean model with thermodynamics. {\em Rev. Mex. F\'is.\/}
  {\bf 41}, 117--135.

\bibitem[Ripa(1996{\natexlab{{\em b\/}}})]{Ripa96}
{\sc Ripa, P.} 1996{\natexlab{{\em b\/}}} Linear waves in a one-layer ocean
  model with thermodynamics. {\em J. Geophys. Res.\/} {\bf 101}, 1233--1245.

\bibitem[R{\o}ed(1997)]{Roed97}
{\sc R{\o}ed, L.~P.} 1997 Energy diagnostics in a $1\frac{1}{2}$-layer,
  nonisopycnic model. {\em J. Phys. Oceanogr.\/} {\bf 27}, 1472--1476.

\bibitem[R{\o}ed \& Shi(1999)]{roed1999numerical}
{\sc R{\o}ed, L.~P. \& Shi, X.~B.} 1999 A numerical study of the dynamics and
  energetics of cool filaments, jets, and eddies off the {Iberian Peninsula}.
  {\em J. Geophys. Res.\/} {\bf 104}, 29817--29841.

\bibitem[Saito \& Ishioka(2014)]{SaitoIshioka14}
{\sc Saito, I. \& Ishioka, K.} 2014 {Mechanism for the formation of equatorial
  superrotation in forced shallow-water turbulence with Newtonian cooling}.
  {\em J. Atmos. Sci.\/} {doi:10.1175/JAS-D-14-0235.1, in press}.

\bibitem[Schneider \& Liu(2009)]{SchneiderLiu09}
{\sc Schneider, T. \& Liu, J.} 2009 {Formation of jets and equatorial
  superrotation on Jupiter}. {\em J. Atmos. Sci.\/} {\bf 66}, 579--601.

\bibitem[Schopf \& Cane(1983)]{cane1983equatorial}
{\sc Schopf, P.~S. \& Cane, M.~A.} 1983 On equatorial dynamics, mixed layer
  physics and sea surface temperature. {\em J. Phys. 0ceanogr.\/} {\bf 13},
  917--935.

\bibitem[Schubert {\em et~al.\/}(2009)Schubert, Taft \&
  Silvers]{schuberttaftsilvers09}
{\sc Schubert, W.~H., Taft, R.~K. \& Silvers, L.~G.} 2009 Shallow water
  quasi-geostrophic theory on the sphere. {\em J. Adv. Model. Earth Syst.\/}
  {\bf 1}, 2.

\bibitem[Scott \& Polvani(2007)]{ScottPolvani07}
{\sc Scott, R.~K. \& Polvani, L.~M.} 2007 Forced-dissipative shallow-water
  turbulence on the sphere and the atmospheric circulation of the giant
  planets. {\em J. Atmos. Sci.\/} {\bf 64}, 3158--3176.

\bibitem[Scott \& Polvani(2008)]{ScottPolvani08}
{\sc Scott, R.~K. \& Polvani, L.~M.} 2008 Equatorial superrotation in shallow
  atmospheres. {\em Geophys. Res. Lett.\/} {\bf 35}, L24202.

\bibitem[Shepherd(1993)]{Shepherd93}
{\sc Shepherd, T.~G.} 1993 A unified theory of available potential energy. {\em
  Atmosphere-Ocean\/} {\bf 31}, 1--26.

\bibitem[Showman(2007)]{showman2007numerical}
{\sc Showman, A.~P.} 2007 Numerical simulations of forced shallow-water
  turbulence: {Effects} of moist convection on the large-scale circulation of
  {Jupiter} and {Saturn}. {\em J. Atmos. Sci.\/} {\bf 64}, 3132--3157.

\bibitem[Srinivasan \& Young(2012)]{SrinivasanYoung12}
{\sc Srinivasan, K. \& Young, W.~R.} 2012 Zonostrophic instability. {\em J.
  Atmos. Sci.\/} {\bf 69}, 1633--1656.

\bibitem[Srinivasan \& Young(2014)]{SrinivasanYoung14}
{\sc Srinivasan, K. \& Young, W.~R.} 2014 Reynolds stress and eddy diffusivity
  of $\beta$-plane shear flows. {\em J. Atmos. Sci.\/} {\bf 71}, 2169--2185.

\bibitem[Thompson(1971)]{Thompson71o}
{\sc Thompson, R. O. R.~Y.} 1971 {Why there is an intense Eastward current in
  the North Atlantic but not in the South Atlantic}. {\em J. Phys. Oceanogr.\/}
  {\bf 1}, 235--237.

\bibitem[Thuburn \& Lagneau(1999)]{ThuburnLagneau99}
{\sc Thuburn, J. \& Lagneau, V.} 1999 Eulerian mean, contour integral, and
  finite-amplitude wave activity diagnostics applied to a single-layer model of
  the winter stratosphere. {\em J. Atmos. Sci.\/} {\bf 56}, 689--710.

\bibitem[Vallis(2006)]{vallis2006atmospheric}
{\sc Vallis, G.~K.} 2006 {\em Atmospheric and {Oceanic} {Fluid} {Dynamics}\/}.
  Cambridge University Press.

\bibitem[Vasavada \& Showman(2005)]{VasavadaShowman05}
{\sc Vasavada, A.~R. \& Showman, A.~P.} 2005 {Jovian atmospheric dynamics: an
  update after Galileo and Cassini}. {\em Rep. Prog. Phys.\/} {\bf 68},
  1935--1996.

\bibitem[Verkley(2009)]{verkley09}
{\sc Verkley, W. T.~M.} 2009 A balanced approximation of the one-layer
  shallow-water equations on a sphere. {\em J. Atmos. Sci.\/} {\bf 66},
  1735--1748.

\bibitem[Walterscheid {\em et~al.\/}(2000)Walterscheid, Brinkman \&
  Schubert]{Walterscheid2000}
{\sc Walterscheid, R.~L., Brinkman, D.~G. \& Schubert, G.} 2000 {Wave
  disturbances from the comet SL--9 impacts into Jupiter's atmosphere}. {\em
  Icarus\/} {\bf 145}, 140--146.

\bibitem[Warneford(2014)]{ESWthesis}
{\sc Warneford, E.~S.} 2014 {\em The thermal shallow water equations, their
  quasi-geostrophic limit, and equatorial super-rotation in Jovian
  atmospheres\/}. DPhil thesis, University of Oxford,
  \url{http://ora.ox.ac.uk/objects/uuid:6604fcac-afe6-4abe-8a6f-6a09de4f933f}.

\bibitem[Warneford \& Dellar(2013)]{WarnefordDellarQGTSW}
{\sc Warneford, E.~S. \& Dellar, P.~J.} 2013 The {quasi-geostrophic} theory of
  the thermal shallow water equations. {\em J. Fluid Mech.\/} {\bf 723},
  374--403.

\bibitem[Warneford \& Dellar(2014)]{WarnefordDellar14tsw}
{\sc Warneford, E.~S. \& Dellar, P.~J.} 2014 {Thermal shallow water models of
  geostrophic turbulence in Jovian atmospheres}. {\em Phys. Fluids\/} {\bf 26},
  016603.

\bibitem[Williams(1978)]{Williams78}
{\sc Williams, G.~P.} 1978 {Planetary circulations: 1. Barotropic
  representation of Jovian and terrestrial turbulence}. {\em J. Atmos. Sci.\/}
  {\bf 35}, 1399--1426.

\bibitem[Williams \& Yamagata(1984)]{williams1984geostrophic}
{\sc Williams, G.~P. \& Yamagata, T.} 1984 {Geostrophic regimes, intermediate
  solitary vortices and Jovian eddies}. {\em J. Atmos. Sci.\/} {\bf 41},
  453--478.

\bibitem[Yamagata \& Philander(1985)]{YamagataPhilander85}
{\sc Yamagata, T. \& Philander, S.} 1985 The role of damped equatorial waves in
  the oceanic response to winds. {\em J. Oceanogr. Soc. Japan\/} {\bf 41},
  345--357.

\end{thebibliography}
\end{document}